\def\year{2018}\relax
\documentclass[letterpaper]{article} %DO NOT CHANGE THIS
\usepackage{aaai18}  %Required
\usepackage{times}  %Required
\usepackage{helvet}  %Required
\usepackage{courier}  %Required
\usepackage{url}  %Required
\usepackage{graphicx}  %Required
\usepackage{mdframed}
\nocopyright
 
\begin{document}
%\title{Online Abuse of UK MPs from 2015 to 2019: Coalition Choler and Brexit Bile}
\title{Online Abuse of UK MPs from 2015 to 2019: Working Paper}

\author{Mark A. Greenwood, Mehmet E. Bakir, Genevieve Gorrell,\\
\Large{\textbf{Xingyi Song, Ian Roberts and Kalina Bontcheva}}\\
\Large{University of Sheffield, UK}\\
\tt{\{m.a.greenwood,m.e.bakir,g.gorrell,}\\
\tt{x.song,i.roberts,k.bontcheva\}@sheffield.ac.uk}
}

\maketitle
\begin{abstract}
   We extend previous work about general election-related abuse of UK MPs with two new time periods, one in late 2018 and the other in early 2019, allowing previous observations to be extended to new data and the impact of key stages in the UK withdrawal from the European Union on patterns of abuse to be explored. The topics that draw abuse evolve over the four time periods are reviewed, with topics relevant to the Brexit debate and campaign tone showing a varying pattern as events unfold, and a suggestion of a ``bubble'' of topics emphasized in the run-up to the highly Brexit-focused 2017 general election. Brexit stance shows a variable relationship with abuse received. We find, as previously, that in quantitative terms, Conservatives and male politicians receive more abuse. Gender difference remains significant even when accounting for prominence, as gauged from Google Trends data, but prominence, or other factors related to being in power, as well as gender, likely account for the difference associated with party membership. No clear relationship between ethnicity and abuse is found in what remains a very small sample (BAME and mixed heritage MPs). Differences are found in the choice of abuse terms levelled at female vs. male MPs.
\end{abstract}

%%%%%%%%%%%%%%%%%%%%%%%%%%%%%%%%%%%%%%%%%%%%%%%%%%%%%%%%%%%%%%%%%%%%%%%

\section{Introduction}
\label{sec:intro}

As social media become an inescapable part our lives, the potential for harm is increasingly a focus for attention and concern. Among apparent harms are incivility and threatening behaviour, which proliferate in the anonymized and consequence-free environment of the internet. Where threats and abuse are levelled at our elected representatives, there is potential for negative impact on democracy. MPs are increasingly raising concerns about the threats and abuse they are subjected to, and a third of women MPs have considered quitting as a result of online and verbal abuse received.\footnote{\small{\url{https://www.bbc.co.uk/news/uk-politics-38736729}}}

In 2015 a majority Conservative government succeeded the previous Conservative/Liberal Democrat coalition in the UK. One of the promises made by the Conservatives was that they would hold a referendum on UK European Union membership. This was duly held, and in June 2016, 52\% of voters said that they wanted the UK to leave the EU. With Theresa May taking on the role of Prime Minister, a further general election was held, in which the Conservatives lost their majority, and the consequences of the referendum in terms of a changing relationship with the European Union are, at the time of writing, unfolding under a minority Conservative government supported, via a confidence and supply arrangement, by the 10 MPs of Northern Ireland's Democratic Unionist party. These events were largely unpredicted, and a climate of division has prevailed throughout.

Against this backdrop, there is a need for a comprehensive and objective quantification of the extent of the issue. Here, we provide such an analysis in the context of the Twitter platform, and cover four separate time periods across five years, making it possible to learn more about how current events affect abuse sent, any suggestion of trends, and what findings remain steady across a range of samples. We focus on tweets using obscene nouns (``cunt'', ``twat'', etc.), racist or otherwise bigoted language and milder insults (``you idiot'', ``coward''). In this way, we define ``abuse'' broadly; ``hate speech'', where religious, racial or gender groups are denigrated, would be included in this.

\vspace{2mm}
\begin{mdframed}
\subsection{A note about gender}
Quantitative findings such as those in our work do not address the complex culture surrounding what it means to be female, male or otherwise in our society. Work from the Pew Research Center (\citeyear{pew2014}), as well as expanding on similar findings about female vs. male online abuse levels in a random sample, notes that women were ``more likely to be upset'' by abuse received. Indeed, the meaning of threats and abuse for any particular demographic varies a great deal depending on factors such as history of oppression, physical vulnerability and the security of their social standing. As Chemaly comments in her piece for Time, ``women take online harassment more seriously not because we are hysterics, but because we reasonably have to''~\cite{chemaly2014}. This work presents, therefore, only a part the story. Pew also find that younger women, in the 18-24 age range, are disproportionately targeted; a group not represented in our sample of politicians.\\
\end{mdframed}
\vspace{2mm}

But we do not limit our analysis to hate speech. We included all manner of personal attacks and threats, but not obscene language (e.g. ``fucking'', ``bloody'') as it is less likely to be targeted at the politician personally. In this work, we build on previous work~\cite{gorrell2018twits} by adding two further time periods; one from late 2018 and a further period from early 2019. These time periods offer fruitful grounds for exploration on two counts. Firstly, they provide additional data on the same parliament as was studied in 2017. Data from 2015 and 2017 focus on quite a different set of prominent politicians. In introducing more data from the same parliament as the 2017 set, we have more grounds for observing which effects arise from those particular individuals and which transcend them. Furthermore, the late 2018 and early 2019 datasets come at a time when the spotlight in on the process of negotiating the terms on which the UK proposes to exit the European Union. So Brexit forms a particular focus for interest in this work.

Our previous work~\cite{gorrell2018twits} offered an analysis of Twitter users who send abuse; for example, how they compare to other Twitter users in terms of account age, followers and so forth. We found important differences suggesting different profiles of those who send abuse. This has not formed part of the focus here however. Instead of focusing on who sends abuse, in this work we introduce a new focus in a final section analyzing how abuse terms are used. This is partly in response to a need to expand further on gender, given that our finding that men receive more abuse in quantitative terms runs contrary to the common perception. Please see the box ``A note about gender'' on page 1 for more on this. In this work, we also no longer consider threat terms, but focus just on abusive language.

%%%%%%%%%%%%%%%%%%%%%%%%%%%%%%%%%%%%%%%%%

\section{Related Work}
\label{sec:related}

Whilst online fora have attracted much attention as a way of exploring political
dynamics~\cite{nulty2016social,kaczmirek2013social,colleoni2014echo,weber2012mining,conover2011political,gonzalez2010emotional},
and the effect of abuse and incivility in these contexts has been
explored~\cite{vargo2017socioeconomic,rusel2017bringing,gervais2015incivility,hwang2008does},
little work exists regarding the abusive and intimidating ways people address politicians online; a trend that has worrying implications for democracy. Theocharis et al~\shortcite{theocharis2016bad} collected
tweets centred around candidates for the European Parliament election in 2014 from Spain, Germany, the United Kingdom and France posted in the month surrounding the election. They find that the extent of the abuse and harrassment a politician is subject to correlates with their engagement with the medium. Ward and McLoughlin~\citeyear{ward2017turds} find similar results to ours, for example regarding the greater abuse received by male MPs, in a two and a half month period running from late 2016 to early 2017; we contribute with a more in-depth exploration of how prominence relates to abuse received, and by studying four time periods over five years. They consider hate speech/gendered slurs separately, however, and find that women receive more of these. They find that most Twitter abuse takes the form of a reply.

The subject of online abuse of women politicians and journalists has also been taken up in a campaign by Amnesty International. Stambolieva's work for Amnesty~\cite{stambolieva2017methodology} studied only women MPs, though has been interpreted as indicating that women receive more abuse than men.\footnote{\small{\url{http://tinyurl.com/leaf-blog-wom-harrass}}; see also correction at \small{\url{http://tinyurl.com/guardian-abbott-most-ab}}} It found that Diane Abbott received almost half of abusive tweets targeted at women MPs in the six weeks preceding the 2017 general election period, far more than any other woman MP, a finding which was then reported in the press.\footnote{\small{\url{http://tinyurl.com/guardian-may-abuse}}}\footnote{\small{\url{http://tinyurl.com/guardian-abbott-most-ab}}} However we find that Theresa May received almost eight times as many abusive tweets as Diane Abbott in the month preceding the election, and more abuse as a proportion of total tweets received. Recent work in association with Amnesty continues the programme,\footnote{\small{\url{http://tinyurl.com/amnesty-unsocial-media}}}\footnote{\small{\url{http://tinyurl.com/amnesty-toxic-place}}} again studying only women. Recently, the programme has crowdsourced a large corpus of abuse against women MPs and journalists~\cite{delisle2019large}.\footnote{\small{\url{http://tinyurl.com/amnesty-crowdsourced}}}\footnote{\small{\url{http://tinyurl.com/amnesty-troll-findings}}}

%Removed citations due to lack of space: miro2016cyber (1 citation), silva2016analyzing (20 citations), awan2017will (12 citations), blackburn2014stfu 44,kwak2015exploring 30
%Keeping perry2009cyberhate 59,coe2014online 151,cheng2015antisocial 96
A larger body of work has looked at hatred on social media more
generally~\cite{perry2009cyberhate,coe2014online,cheng2015antisocial}. Williams~\cite{williams2015cyberhate}
and Burnap~\cite{burnap2015cyber} present work demonstrating the potential of Twitter for
evidencing social models of online hate crime that could support
prediction, as well as exploring how attitudes co-evolve with events
to determine their impact. Silva et
al~\shortcite{silva2016analyzing} use natural language processing (NLP) to
identify the groups targeted for hatred on Twitter and Whisper.
Munger presents intriguing work on automated (bot) social sanctions, (e.g.~\citeauthor{munger2017tweetment}~\citeyear{munger2017tweetment}).

A surge of recent interest aims to detect abuse, hate speech and toxicity automatically, resulting in increasing availability of training data and workshops focused on the topic~\cite{burnap2015cyber,nobata2016abusive,chen2012detecting,dinakar2012common,wulczyn2017ex,bretschneider2017detecting,nobata2016abusive,waseem2016hateful}.\footnote{\small{\url{http://tinyurl.com/alw-workshop-2017}}}\footnote{\small{\url{http://tinyurl.com/alw-workshop-2018}}}\footnote{\small{\url{http://tinyurl.com/alw-stack-overflow}}} Schmidt and
Wiegand~\shortcite{schmidt2017survey} provide a review of prior work
and methods, as do Fortuna and Nunes~\shortcite{fortuna2018survey}. Reported performances vary widely depending on the specifics of the task. In very recent times attention has begun to turn to the issue of bias in abuse classification. Unintended bias, for example being more likely to label a text as abusive if it was penned by a particular demographic because that bias was present in the training data, has been highlighted as an issue in Google's ``Perspective'' toxicity scorer,\footnote{\small{\url{http://tinyurl.com/jigsaw-unintend-bias}}} which they have recognized, and the overcoming of which is formulated as an explicit objective in their new competition.\footnote{\small{\url{http://tinyurl.com/jigsaw-kaggle-toxic}}} Whilst unintended bias has been the subject of much research in recent years with regards to making predictive systems that don't discriminate, it has only just begun to be taken up within abuse classification~\cite{park2018reducing}.

In the next section, we describe our data collection methodology. We then present our results, beginning with an analysis of who receives the abuse, before moving on to the topics that are most likely to trigger abusive replies. We then discuss some differential patterns in term use, before concluding.

%%%%%%%%%%%%%%%%%%%%%%%%%%%%%%%%%%%%%%%%%

\section{Methods}

\subsection{Data Collection}
\label{sec:data}

%\cite{Maynard17a}
%https://gist.github.com/greenwoodma/
The corpora were created by downloading tweets in real-time using Twitter's streaming API. The data collection focused on Twitter accounts of MPs, candidates, and official party accounts. We obtained a list of all current MPs\footnote{From a list made publicly available by BBC News Labs,
which we cleaned and verified} and all currently known election candidates\footnote{List of candidates obtained from \small{\url{https://yournextmp.com}}} (at that time) who had Twitter accounts.

We used the API to follow the accounts of all MPs over the period of interest. This means we collected all the tweets sent by each MP, any replies to those tweets, and any retweets either made by the MP or of the MPs own tweets. Note that this approach does not collect all tweets which an MP would see in their timeline as it does not include those in which they are just mentioned. We took this approach as the analysis results are more reliable  due to the fact that replies are directed at the politician who authored the tweet, and thus, any abusive language is more likely to be directed at them.

Data were of a low enough volume not to be constrained by Twitter rate limits. Corpus statistics are given in table~\ref{tab:corpus}, and are separated out into all politicians studied and just those who were then elected as MPs. The following time periods define the data used in this work:

\begin{itemize}

    \item \textbf{2015}: 7 April 2015 00:00 am - 7 May 2015 22:00 pm. This is the month before the 2015 general election, finishing at the same time as the polls closed; hence the 10pm cutoff;

    \item \textbf{2017}: 8 May 2017 00:00 am  -  8 June 2017 22:00 pm. This is the month before the 2017 snap general election, finishing at the same time as the polls closed; hence the 10pm cutoff;

    \item \textbf{2018}: 21 Nov 2018 - 21 Dec 2018. This period includes the 11th of December when there was planned to be a meaningful vote on Brexit which was pulled at the last minute on the 10th of December;

    \item \textbf{2019}: 7 Jan 2019 - 6 Feb 2019. This period includes the meaningful vote in which the government suffered the largest ever defeat in the house of commons.

\end{itemize}

\begin{table*}
\begin{center}
  \begin{tabular}{l|l|l|l|l|l|l|l}
\textbf{Date} & \textbf{\# MPs} & \textbf{Tw. sent by MPs} & \textbf{Replies to MPs} & \textbf{Abusive Repl.} & \textbf{\% Abusive} & \textbf{Retweets of MPs} & \textbf{Avg \# RTs}\\
\hline
2015 & 474 & 59,143 & 267,867 & 7,960 & 2.97\% & 412,520 & 6.97\\
2017 & 541 & 49,984 & 614,691 & 20,477 & 3.33\% & 1,706,102 & 34.13\\
2018 & 501 & 37,405 & 1,013,538 & 30,871 & 3.05\% & 1,795,952 & 48.01\\
2019 & 504 & 36,210 & 939,982 & 26,854 & 2.86\% & 1,735,576 & 47.93\\
 \end{tabular}
\caption{Corpus statistics}
\label{tab:corpus}
\end{center}
\end{table*}

\subsection{Abuse Classification}

In order to identify abusive language, the politicians it is targeted at, and the topics in the politician's original tweet that tend to trigger abusive replies, we use a set of NLP
tools, combined into a semantic analysis pipeline. It includes, among other things, a component for MP and candidate recognition, which detects mentions of MPs and election candidates in the tweet and pulls in information about them from DBpedia. Topic detection finds mentions in the text of political topics (e.g. environment, immigration) and subtopics (e.g. fossil fuels). The list of topics was derived from the set of topics used to categorise documents on the gov.uk website\footnote{e.g. \small{\url{https://www.gov.uk/government/policies}}}, first
seeded manually and then extended semi-automatically to include
related terms and morphological variants using
TermRaider\footnote{\small{\url{https://gate.ac.uk/projects/arcomem/TermRaider.html}}}, resulting in a total of 940 terms across 51 topics.
This methodology is presented in more detail in Maynard et
al~\citeyear{Maynard17a}.
We also perform hashtag tokenization, in order to find abuse and threat terms that otherwise would be missed. In this way, for example, abusive language is found in the hashtag ``\#killthewitch''.
%, which was aimed at Theresa May in the
%tweet (paraphrased): \textit{``@theresa\_may If you legalise
%  \#foxhunting I will shoot an arrow at your face. \#promises
%  \#killthewitch''}
%https://twitter.com/Powderedcows/status/862087943820857344.

A dictionary-based approach was used to detect abusive language in tweets. An abusive tweet is considered to be one containing one or more abusive terms from the vocabulary list.\footnote{Warning; strong language and offensive slurs: \small{\url{http://staffwww.dcs.shef.ac.uk/people/G.Gorrell/publications-
materials/abuse-terms.txt}}} This contained 388 abusive terms or short phrases in British and American English, comprising mostly an extensive collection of insults. Racist and homophobic terms are included as well as various terms that denigrate a person's appearance or intelligence. Short phrases might include ``you'' to increase precision, for example ``you cock'' rather than ``cock''.

Data from Kaggle's 2012 challenge, ``Detecting Insults in Social
Commentary''\footnote{\small{\url{https://www.kaggle.com/c/detecting-insults-in-social-commentary/data}}}, was used to evaluate the success of the approach. The training set was used to tune the terms included. On the test set, our approach was shown to have an accuracy of 0.80 (Cohen's Kappa: 0.41), with a precision of 0.70, a recall of 0.42 and an F1 of 0.52. In practice, performance is likely to be slightly better than this, since the Kaggle corpus is American English.

Data-driven approaches have achieved widespread popularity through their typically superior results, but with the increasing real-world use of prediction systems, concerns have recently grown about their tendency to contain unwanted bias~\cite{hardt2016equality,bolukbasi2016man,caliskan2017semantics}. In the case of abuse detection, an example would be a system that learns from the training data that men receive more abuse, so on unseen data uses indicators of a male target, such as a male name, to classify a text as abusive. Even a simple data-driven approach, if it has access to all the words in the input as features, is likely to learn more from the training data than strictly pertains to the task. In this work, this would be completely unacceptable, as it would make it impossible for us to know if men really did receive more abuse in unseen data, or if the classifier was biased. Subtler unwanted bias might include stereotyping a particular cultural group as more likely to be abusive, so labelling texts containing certain slang as abusive even when they aren't. An abundance of recent work seeks to address this problem (e.g.~\cite{dixon2018measuring,zhao2018gender}), but critics say we are still far from having solved it~\cite{elazar2018adversarial,pleiss2017fairness,gonen2019lipstick}. For complete peace of mind, a rule-based approach offers much more transparency and less scope for unwanted bias.

However, in order to give a baseline demonstrating the adequacy of abuse classification performance, we also evaluated a convolutional neural net (CNN) approach on the same (Kaggle) data. Our model contains three major parts. We first encode the input words with a pre-trained 100 dimension Glove twitter embedding \cite{pennington2014glove}. Then we apply a CNN network \cite{Kim14f} (kernel size 3, 4, 5, and 100 output dimensions), with batch normalization (momentum=0.1) to extract the local features of the input text. Finally, our output layer applies a sigmoid function for binary classification.

We obtained a result somewhat superior to results obtained by others on that dataset when the competition ran in 2012,\footnote{\small{\url{https://www.kaggle.com/c/detecting-insults-in-social-commentary/leaderboard}}} with an AUROC of 0.8662. Our rule-based system cannot easily be compared directly, as AUROC does not make sense for such an approach. The result was fractionally better than the rule-based approach, with an accuracy of 0.8243 and Cohen’s Kappa of 0.4765. We also measured the positive class recall (0.4603), precision (0.7780) and F1-measure (0.5784). Given that this system was trained and tested on data from the same source, which was longer social media posts in American English, the result cannot be expected to carry over undiminished to our British English Twitter data. We therefore concluded that our gazetteer system was acceptable in performance and also more suitable through its resistance to unwanted bias.

Inspection of errors in the rule-based system reveals that performance is diminished by coarse language in non-abusive texts and polite language in abusive texts, but nothing that raised concerns for the results presented here.

\subsection{Topic Identification}

Topic analysis of tweets is based on reusing the methods from the Nesta Political Futures Tracker project (PFT).~\footnote{\small{\url{https://www.nesta.org.uk/blog/introducing-the-political-futures-tracker/}}} The PFT project focused on real-time analysis of tweets in the run-up to the May 2015 General Election.

The set of topics we adopted were based around the UK government’s high-level ``policy areas'',\footnote{\small{\url{https://www.gov.uk/government/topics}}} such as public health, immigration, and law, although we have extended this list to include other highly relevant topics including Brexit. For each of these, we then associated sets of keywords (e.g. for public health: NHS, nursing, doctors), which were discovered automatically from the party manifestos and UK election tweets, then post-edited and extended manually.

\subsection{Analysis}

When we look at the abuse received by MPs there are both many ways to subdivide the MPs into groups as well as different ways of then analysing the data within the groupings. Many of the remaining sections of this document focus on a specific partitioning of the MPs into different groups, be that by party, gender, or a political stance etc. Within each of the following sections the level of abuse can, and often is, reported in at least two different ways. One way of analysing the data for a group of MPs is simply to add together all the abusive tweets received by MPs in that group and report that as a percentage of the total number of replies to the same MPs. This gives a single headline figure such as ``3.8\% of replies to Conservative MPs in 2015 were abusive'' but completely ignores the possible wide variation amongst members of the group. An alternative is to determine the abuse level per MP and then look at the mean value (and standard deviation) which, for example, shows that a Conservative MP in 2015 is likely to see 0.775\% of their replies being abusive.

Both of these approaches, while giving very different numbers, are equally valid ways of looking at the data, but care needs to be taken when interpreting the results to understand how the analysis was performed.

%%%%%%%%%%%%%%%%%%%%%%%%%%%%%%%%%%%%%%%%%%%%%%%%%%%%%%%%%%%%%%%%%%%%%%%

\section{Findings}

Findings are presented under three headings. First, who is receiving the abuse, in terms of gender, ethnicity and political position. Secondly, we investigate the topics that arise in abusive tweets particularly. Finally we look at particular abuse words, how they are used differently and trends over time.

\subsection{Who is receiving the abuse?}

We begin with an overview of the data, that allows an overall sense of the targets of abuse and the magnitude of the issue to be gained, in the context of the political climate. We then review individual characteristics and their relationship to abuse under four headings; gender, ethnicity, party and Brexit stance, before bringing data together in a combined model.

\subsubsection{Overview}
%BY WHICH WE MEAN, SUNBURSTS

\begin{figure}[t]
\includegraphics[width=\columnwidth]{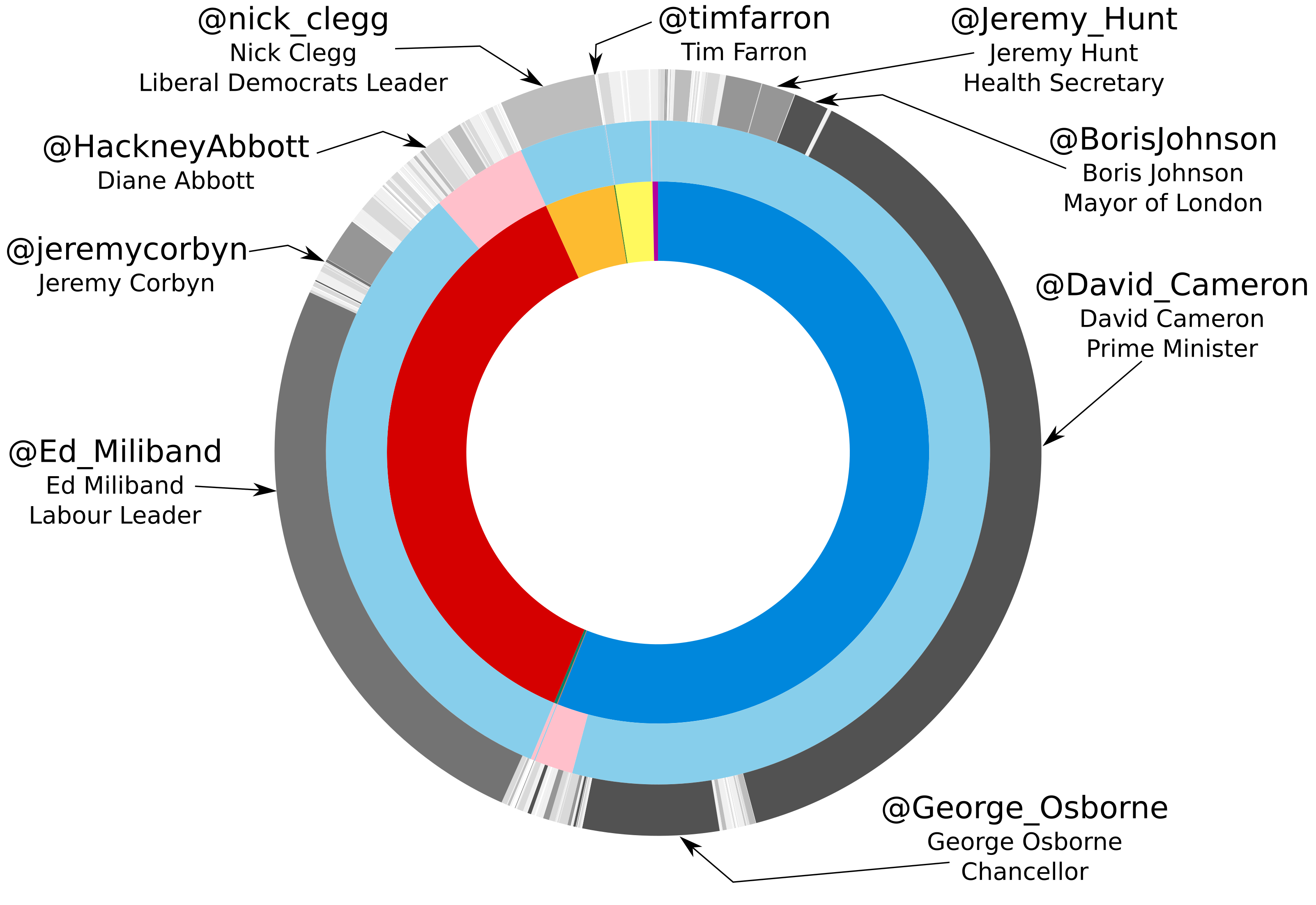}
\caption{Abuse per MP in 2015}
\label{fig:abuse2015Proportion}
\end{figure}

\begin{figure}[t]
\includegraphics[width=\columnwidth]{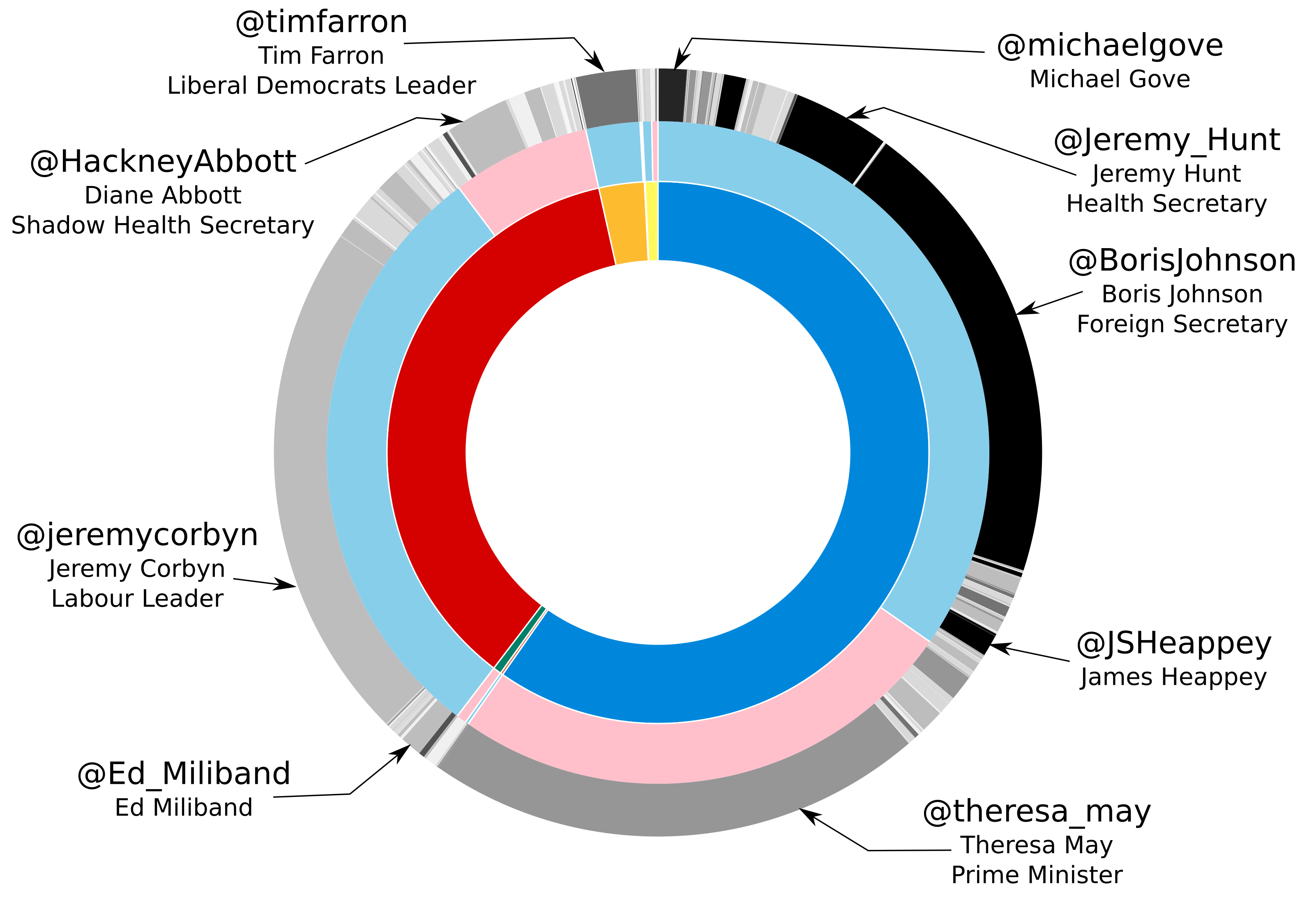}
\caption{Abuse per MP in 2017}
\label{fig:abuse2017Proportion}
\end{figure}

\begin{figure}[t]
\includegraphics[width=0.89\columnwidth]{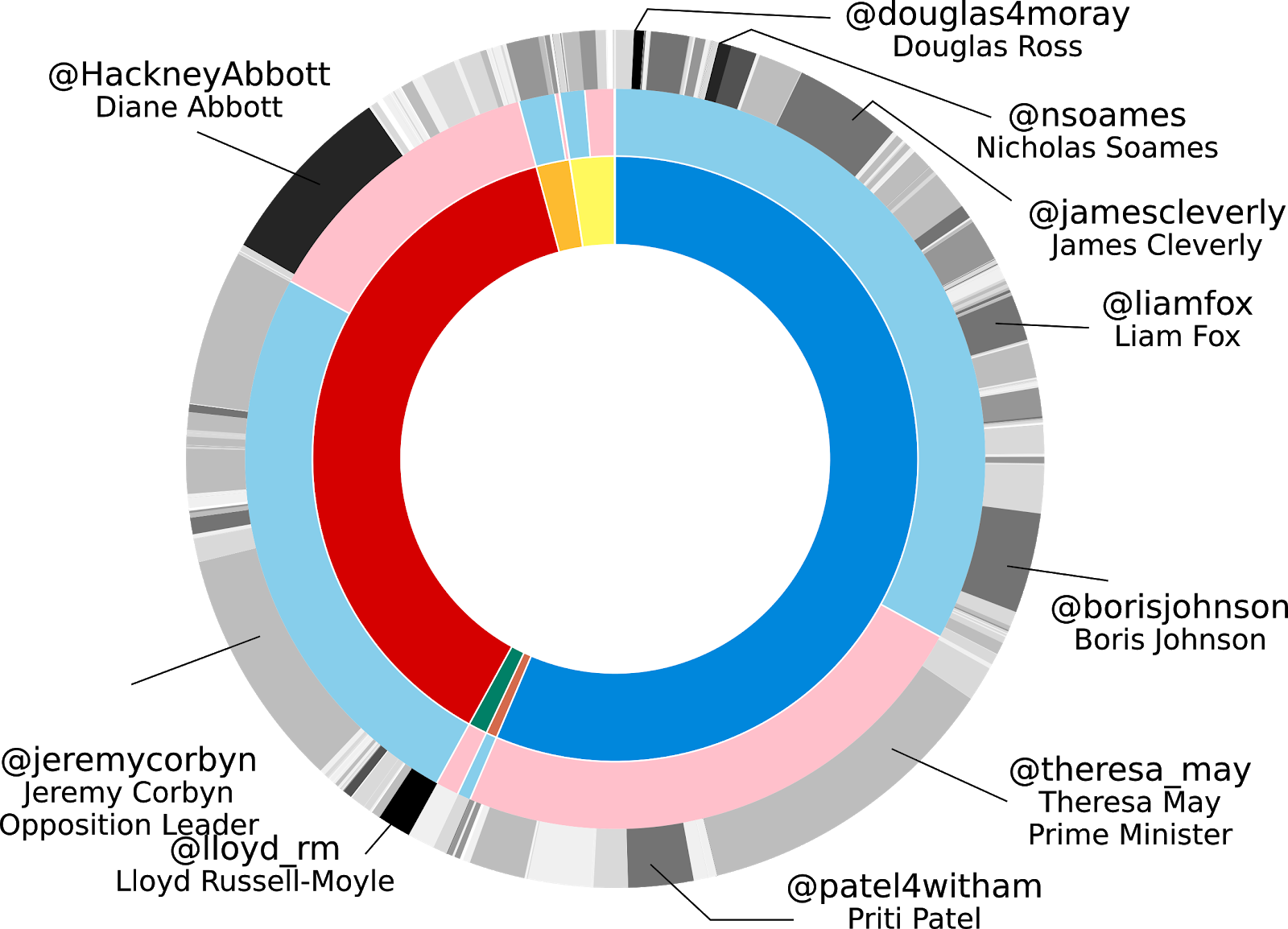}
\caption{Abuse per MP in 2018}
\label{fig:abuse2018Proportion}
\end{figure}

\begin{figure}[t]
\includegraphics[width=0.93\columnwidth]{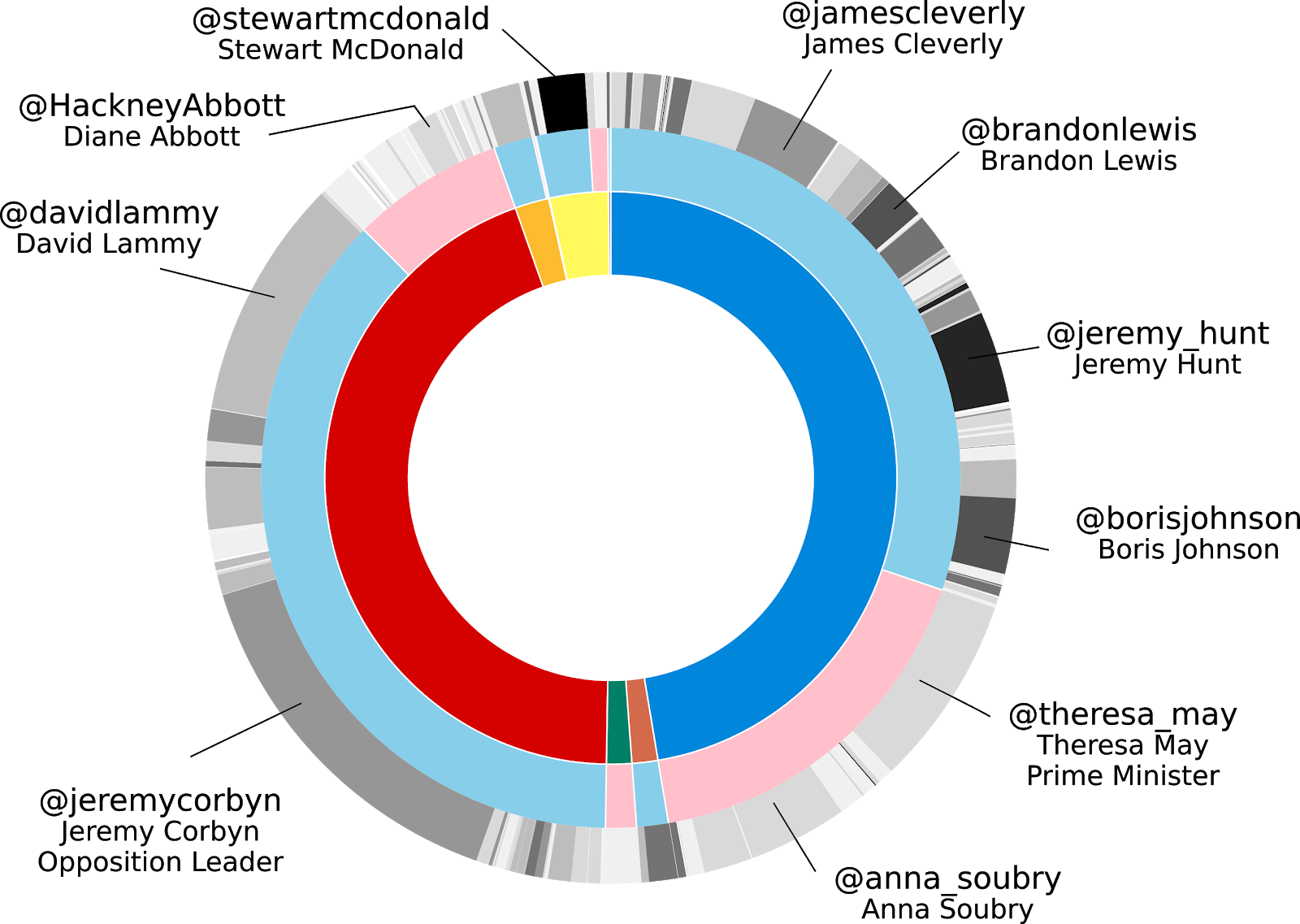}
\caption{Abuse per MP in 2019}
\label{fig:abuse2019Proportion}
\end{figure}

Figures \ref{fig:abuse2015Proportion}, \ref{fig:abuse2017Proportion}, \ref{fig:abuse2018Proportion} and
\ref{fig:abuse2019Proportion} are also available online in interactive
form,\footnote{\small{\url{http://demos.gate.ac.uk/politics/itv/sunburst.html}}} allowing the viewer to explore the abuse received by every MP with an active Twitter account in each of the four time periods studied. The outer ring represents MPs, and the size of each outer segment is
determined by the number of abusive replies each receives. Working inwards you can then see the MP's gender (pink for female and blue for male) and party affiliation (blue for Conservative, red for Labour, yellow for the Scottish National Party -- see the interactive graph to explore the smaller parties). The darkness of each MP segment denotes the percentage of the replies they receive which are abusive. This ranges from white, which represents up to 1\%, through to black, which represents 8\% and above. So an MP with a wide, light-coloured segment tends not to receive a lot of abusive tweets, but because they receive so many tweets this adds up to a lot of abusive tweets. An MP with a narrow, dark segment doesn't receive many tweets, but when they do they are disproportionately likely to be abusive.

In the pre-election periods we see a lot of abusive attention aimed at party leaders. In the 2018 and 2019 periods, abuse distributes itself more evenly. This is most likely because in pre-election periods people are focused on expressing their opinions about the future leadership of the country.

In 2015, shown in figure~\ref{fig:abuse2015Proportion}, abuse was overwhelmingly aimed at the Conservative party, with the Labour party seeing just less than half the amount of the Conservatives. By 2019, shown in figure~\ref{fig:abuse2019Proportion}, the abuse is almost equally split between the two main parties (both receive roughly the same volume and percentage of abuse). We also see an increase in the volume of abuse aimed at women MPs after 2015, not explained by an increase in numbers (the ``count'' setting of the interactive
graphs\footnote{\small{\url{http://demos.gate.ac.uk/politics/itv/sunburst.html}}} makes this much more apparent) though as we will see below, this appearance arises from the prominence of a small number of women MPs, most notably Theresa May. Averages per individual tell a different story.

The most notable targets for abuse tend to be males and Conservatives. In 2015, a few high-profile male figures drew the great majority of the abuse. Whilst Theresa May gained prominence in 2017, Boris Johnson was the most notable target for abuse, as seen in figure~\ref{fig:abuse2017Proportion}. Generally, Boris Johnson has tended to draw abuse. Jeremy Hunt tends to draw abuse.

In 2015, David Cameron received most abuse by volume and a high level of abuse by percent. In terms of volume, Ed Milliband also received much abuse, but by percent, Boris Johnson and George Osborn received more. In 2015, little abuse was aimed at women MPs. In 2017, party leaders Jeremy Corbyn and Theresa May received much abuse by volume, but neither received strikingly much abuse by percentage. The people that were disproportionately targeted with abuse were Boris Johnson and Jeremy Hunt. James Heappey, whilst not receiving a great deal of abuse by volume was disproportionately targeted by abuse, in the wake of an unpopular comment to a Scottish schoolgirl. Michael Gove was also disproportionately targeted.

Diane Abbott received a particularly high volume and percentage of abuse in the late 2018 period, shown in figure~\ref{fig:abuse2018Proportion}, having not been highly targeted in the previous samples. Ms Abbott made a controversial comment about policing around this time, which drew ire from the right and was picked up in a Breitbart article. This perhaps highlights that abuse can vary with current events, and certain sample periods may not be representative of particular individuals' experience. There is reduced attention on party leaders compared with previous years. Lloyd Russell-Moyle announced his HIV status in this time period. Priti Patel received criticism during this time period for a comment regarding food shortages and Ireland. Several women attract significant volumes of abuse in this period, and the two individuals receiving the highest levels of abuse by percentage are Labour politicians.

Theresa May receives remarkably little abusive attention during the early 2019 period; Jeremy Corbyn receives more abuse by volume and percentage. But generally, in this period the abuse is distributed. David Lammy receives more abuse than Theresa May in this period, following from a race row, and Jeremy Hunt and Stewart McDonald receive the highest percentages. Jeremy Hunt tends to receive high levels of abuse; Stewart McDonald was the target of far-right attention.

Whilst this view is useful in understanding the amount of abuse individual politicians receive in absolute terms, it is less valuable in illustrating the different responses the parties and genders receive overall, because individual effects dominate the picture. Therefore it is important to also see the results per-MP. The online version of these graphs includes a ``count'' view in which each segment of the outer ring represents a single MP. It is evident at a glance that replies to male Conservative MPs are proportionally more abusive, with female Conservative MPs not far behind, an impression that will be further explored statistically below. We take gender, ethnicity, party and Brexit stance in turn, before using structural equation modelling to propose a combined model.
%In taking demographic factors separately, it is understood that we are describing the experience of abuse such an individual is likely to have. We are not commenting on causality. As a hypothetical example, a man may receive more abuse \textit{because} he is in the Conservative party. But descriptively speaking, we might still observe that men are receiving more abuse.

\subsubsection{Gender}

Across all four time periods, male politicians on average receive more abuse than their female peers, as shown in table~\ref{tab:men-women-abuse}. In 2015, the difference between the genders was small with female MPs receiving just over 1 abuse tweet in every 200 responses, compared to the 1.4 per 200 abusive responses received by male MPs, and in fact, the difference was not statistically significant (using a non-parametric Mann-Whitney U-test). Since 2017, however, the difference in abuse between male and female MPs has widened slightly and is enough to show a consistent statistically significant difference.

This analysis compares average abuse received by a female or a male MP, whereas the sunburst diagrams above show results dominated by a few high profile individuals, so gives a more cautious impression. In the section ``Combined Model'' below, we are able to abstract out how high profile an MP is, how popular they are on Twitter (in terms of tweets received) and how engaged they are on Twitter. This model also confirms the findings here. Only male and female gender identities are represented in our sample.

\begin{table}
\begin{center}
\resizebox{\columnwidth}{!}{%
  \begin{tabular}{l|l|l|l}
\textbf{Date} & \textbf{Female} & \textbf{Male} & \textbf{Is significant}\\
\hline
2015 & 0.570 ($\sigma = 0.966$) & 0.746 ($\sigma = 1.144$) & No (p = 0.289)\\
2017 & 0.849 ($\sigma = 1.055$) & 1.464 ($\sigma = 2.406$) & Yes (p\textless0.01)\\
2018 & 1.289 ($\sigma = 2.198$) & 1.832 ($\sigma = 1.815$) & Yes (p\textless0.001)\\
2019 & 1.093 ($\sigma = 1.230$) & 1.601 ($\sigma = 1.772$) & Yes (p\textless0.01)\\
 \end{tabular}
 }
\caption{Abusive tweets per 100 responses (Mean Value)}
\label{tab:men-women-abuse}
\end{center}
\end{table}

\begin{figure*}[t]
\centering
\includegraphics[width=.85\textwidth]{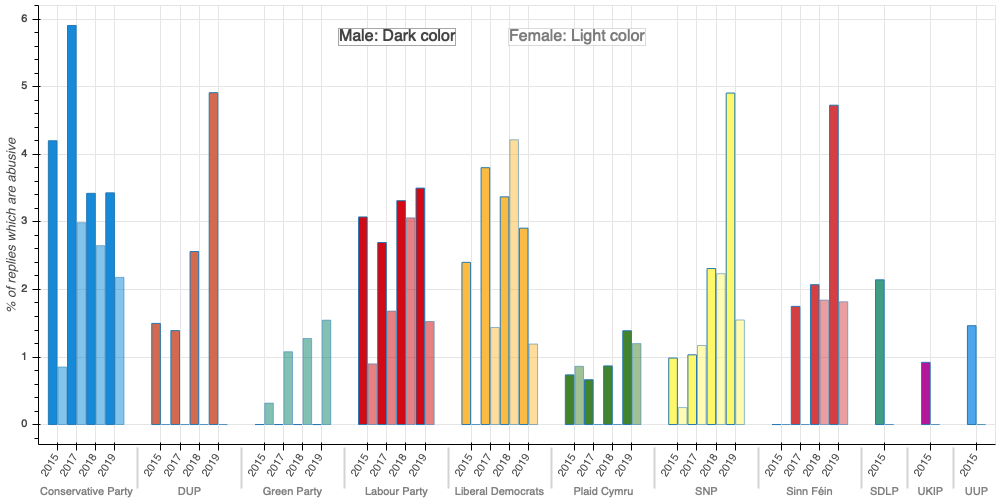}
\caption{Abuse per Gender per Party}
\label{fig:gender-party}
\end{figure*}

It is worth noting that the difference between the genders also appears to hold when viewed at the party level, as shown in figure~\ref{fig:gender-party}. In fact the only occasion when female members of a party received more abuse than their male colleagues was for the Liberal Democrats during the 2018 period. In all other periods male MPs received more abuse than the female MPs of the same party.

\subsubsection{Ethnicity}

We classified each MP as being either white or BAME (Black, Asian, or minority ethnic, including mixed-heritage) and then looked at the abuse each class received. While there were differences in the level of abuse, not only was it not significant, but the class with the higher level of abuse varied across the four time periods, as shown in table~\ref{tab:ethnicity-abuse}. This lack of demonstrable difference could arise from the small percentage (below 10\%) of MPs falling within the BAME/mixed-heritage grouping (34, 46, 42 and 43 for each time period respectively). Other factors such as gender and party membership could also affect the overall figure.

\begin{table}
\begin{center}
\resizebox{\columnwidth}{!}{%
  \begin{tabular}{l|l|l|l}
\textbf{Date} & \textbf{BAME} & \textbf{White} & \textbf{Is significant}\\
\hline
2015 & 0.835 ($\sigma = 1.023$) & 0.678 ($\sigma = 1.097$) & No (p = 0.220)\\
2017 & 1.096 ($\sigma = 1.045$) & 1.274 ($\sigma = 2.146$) & No (p=0.517)\\
2018 & 2.206 ($\sigma = 4.058$) & 1.601 ($\sigma = 1.658$) & No (p=0.886)\\
2019 & 1.352 ($\sigma = 1.477$) & 1.439 ($\sigma = 1.642$) & No (p=0.684)\\
 \end{tabular}
 }
\caption{Abusive tweets per 100 responses (Mean Value)}
\label{tab:ethnicity-abuse}
\end{center}
\end{table}

\subsubsection{Party}

\begin{figure*}[t]
\centering
\includegraphics[width=0.85\textwidth]{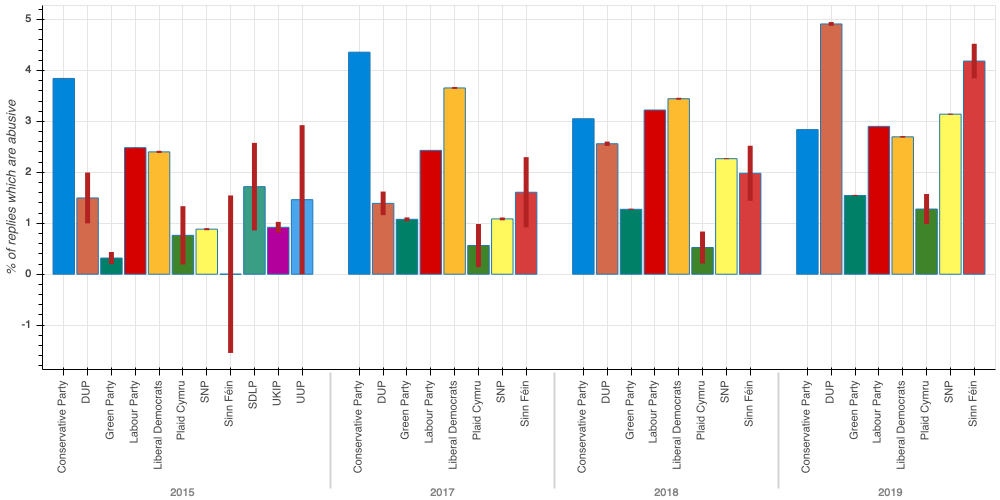}
\caption{Abuse per Party}
\label{fig:abuseParty}
\end{figure*}

Abuse when viewed by party alone is not as clear-cut as some of the other analysis. Firstly outside the main two parties the number of MPs in a party is small (relatively speaking) and not all MPs use Twitter. This has the effect of reducing some parties to just one or two MPs for this analysis, and often they also receive few replies in total. This means that even a single abusive reply can have a huge impact on the percentage of abuse they appear to receive. For example, in 2015 2.4\% of replies to Liberal Democrat MPs were classified as abusive. However, if one looks at the raw data (this is also evident through the per MP sunburst diagrams) this equated to just 335 out of 13,947 tweets. This hardly compares to the roughly similar level of abusive replies (2.5\%) sent to the Labour party over the same period, which was made up of 2,623 abusive tweets out of a total of 106,080 replies. In order to communicate this uncertainty in figure~\ref{fig:abuseParty}, we have added a red bar that indicates how the result would change if that party received three more or three less abusive tweets. It is evident that the impact on the result for the Conservative and Labour parties is insignificant, whereas for smaller parties a few tweets here and there could change the result substantially. Therefore bars with small or invisible ranges should be considered more reliable.

Proceeding, we focus on just the main two parties who have larger numbers of MPs, so we can start to examine the differences for statistical significance. Specifically, if we look at the mean number of abusive tweets an MP could expect to see depending on which party they belong to, then across all four years, Conservative MPs are likely to see more abuse, and this difference is statistically significant in three out of the four periods, as shown in table~\ref{tab:party-abuse}. A possible explanation for this is that, as the party in power, the Conservatives will be held responsible to a greater extent than the opposition for anything that aggrieves people enough to author an abusive tweet.

\begin{table}
\begin{center}
\resizebox{\columnwidth}{!}{%
  \begin{tabular}{l|l|l|l}
  \textbf{Date} & \textbf{Conserv. Party} & \textbf{Labour Party} & \textbf{Is signif.}\\
  \hline
  \hline
2015 & 0.775 ($\sigma=1.282$) & 0.583 ($\sigma=0.759$) & No (p=0.993)\\
  \hline
2017 & 1.580 ($\sigma=2.629$) & 0.913 ($\sigma=1.071$) & Yes (p=0.01)\\
  \hline
2018 & 2.061 ($\sigma=2.242$) & 1.231 ($\sigma=1.526$) & Yes (p\textless0.001)\\
  \hline
2019 & 1.646 ($\sigma=1.889$) & 1.193 ($\sigma=1.152$) & Borderline\\
& & & (p=0.071)\\
 \end{tabular}
 }
\caption{Abusive tweets per 100 responses (Mean Value)}
\label{tab:party-abuse}
\end{center}
\end{table}

It is also worth noting that the periods from 2017 onwards are based on a different set of MPs from those in the 2015 period, due to the snap election held in 2017. This slight difference in the makeup of parliament might explain the difference in abuse levels before and after the 2017 election, as well as the emergence of Brexit and Europe as highly polarizing topics. We will return to the latter point in subsequent sections.

\subsubsection{Brexit Stance}

We now consider Brexit stance as a factor that might influence the amount of abuse received. There are many ways in which we could have labelled MPs as either for or against Brexit, but we choose to focus on how they publicly stated they intended to vote in the referendum. We used a list produced by the Guardian for this classification, although clearly not everyone had a stated position.\footnote{\small{\url{https://www.theguardian.com/politics/ng-interactive/2016/feb/23/how-will-your-mp-vote-in-the-eu-referendum}}}

\begin{table}
\begin{center}
\resizebox{\columnwidth}{!}{%
  \begin{tabular}{l|l|l|l}
\textbf{Date} & \textbf{Remainer} & \textbf{Leaver} & \textbf{Is significant}\\
\hline
2015 & 0.573 ($\sigma = 0.819$) & 0.904 ($\sigma = 1.482$) & No (p=0.493)\\
2017 & 1.234 ($\sigma = 1.640$) & 1.348 ($\sigma = 1.778$) & No (p=0.979)\\
2018 & 1.480 ($\sigma = 1.470$) & 1.758 ($\sigma = 1.730$) & No (p=0.306)\\
2019 & 1.313 ($\sigma = 1.254$) & 1.856 ($\sigma = 2.440$) & No (p=0.714)\\
 \end{tabular}
 }
\caption{Abusive tweets per 100 responses (Mean Value)}
\label{tab:brexit-abuse}
\end{center}
\end{table}

In the post referendum periods we can see that in general abuse aimed at those who wanted to leave the EU and hence back Brexit has continually increased, as shown in table~\ref{tab:brexit-abuse}, whereas the abuse aimed at remainers has stayed at more or less the same level. However, the differences are not statistically significant. Furthermore, we also split the MPs in the same way for 2015, which was before the referendum, and before many MPs would have had a publicly stated position. In that case, leavers still received more abuse than remainers, suggesting that Brexit stance is not causative, but insofar as there is a tendency for leavers to receive more abuse, this arises from other factors, for example gender or party membership. These connections are explored below.

\subsubsection{Combined Model}
\label{sec:combined-model}

Because factors interact, for example women are more likely to be in the Labour party, it can be hard to separate out what is causing particular effects; are women attracting less abuse, or are Labour party members attracting less abuse? Structural equation modelling enables a model relating the various factors to be tested for its fit to the data, allowing the influence of certain factors to be removed from others to give a clearer picture. In our earlier research~\cite{gorrell2018twits}, this allowed us to determine that whilst people more in the public eye (according to their being more searched for on Google) do attract more abuse, that is purely a numbers effect arising from them getting more tweets in general. In fact, people much in the public eye get less abuse than you might expect.

\begin{figure*}[t]
\begin{centering}
\includegraphics[width=.65\textwidth]{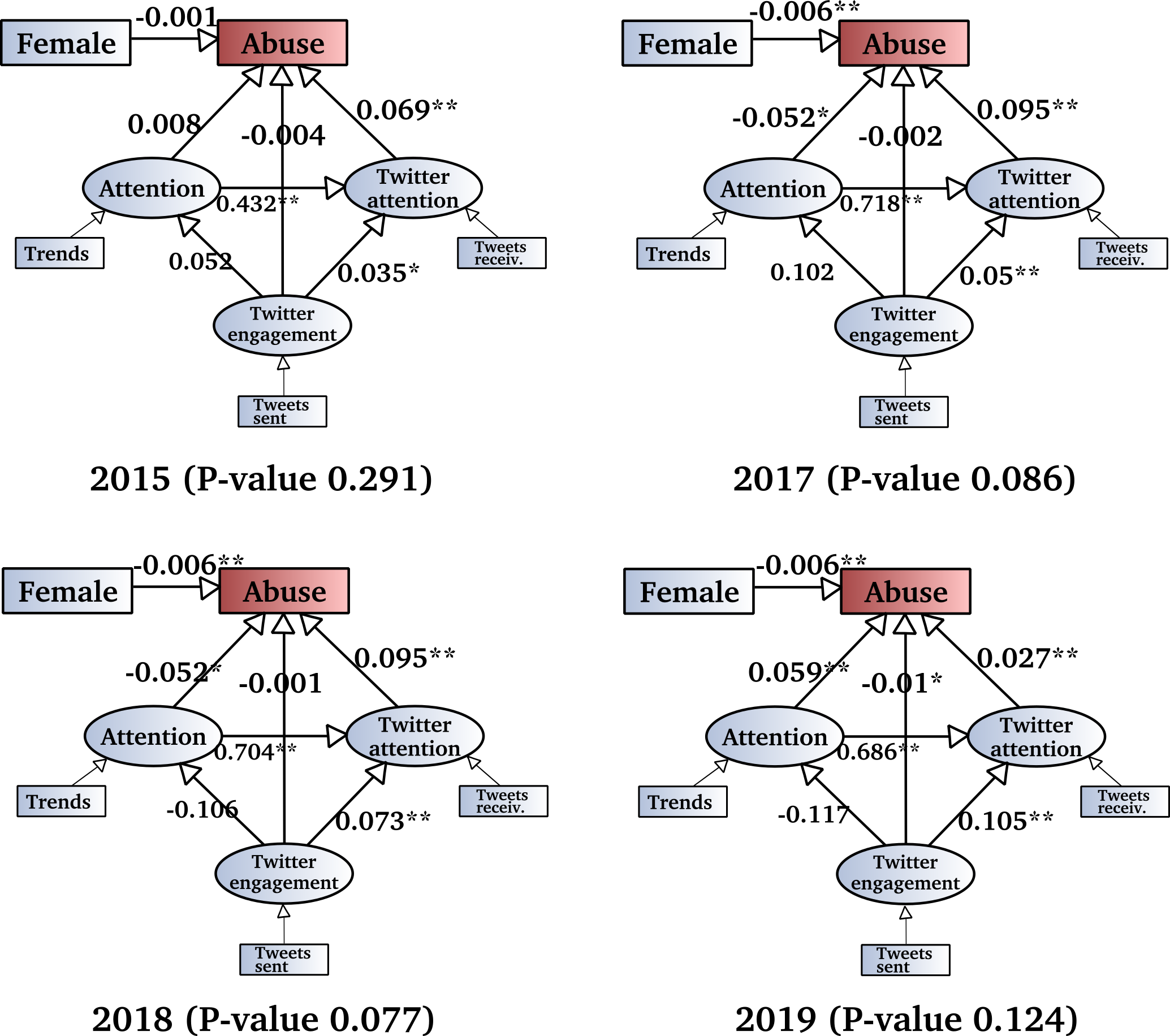}
\caption{SEM models for the four periods}
\label{fig:sem}
\end{centering}
\end{figure*}

Trying the same model across four time periods offers an opportunity to test the robustness of previous findings. Structural equation modelling (see Hox and Bechger~\shortcite{hox2007introduction} for an introduction) was therefore again used to broadly relate three main factors with the amount of abuse
received: prominence, Twitter prominence (which we hypothesise differs from prominence generally) and Twitter engagement. We obtained Google Trends data for the politicians in each of the time periods, and used this variable as a measure of how high-profile that individual is in the minds of the public at the time in question. Search counts were totalled to provide a figure. We used number of tweets sent by that politician as a measure of their Twitter engagement, and tweets received as a measure of how high-profile that person is on Twitter. The model in figure~\ref{fig:sem}, in addition to proposing
that the amount of abuse received follows from these three main factors, also hypothesises that the amount of attention a person receives on Twitter is related to their prominence more generally, and that their engagement with Twitter might get them more attention, both on Twitter and beyond it. It is unavoidably only a partial attempt to describe why a person receives the abuse they do, since it is hard to capture factors specific to that person, such as any recent allegations concerning them, in a measure. The model was fitted using Lavaan.\footnote{\small{\url{http://lavaan.ugent.be/}}}

As previously, we attempted to build a model that includes the personal factors of gender, ethnicity, party and Brexit stance. However, a satisfactory model could not be converged. The only result that remains solid across all datasets is similar to our previous model but includes \textit{only gender} as an influential personal factor on abuse received; not ethnicity, party membership or Brexit stance. This may suggest that an effect due to party membership is lost where prominence is included as a factor. However, the fact that a similar model fits across as many as four separate time periods lends support to the main findings in the earlier work.

Whilst it wasn't possible to fit a model explaining e.g. Brexit stance in terms of gender, ethnicity and party membership, from the available variables, the reader may find it informative to learn the correlations between these variables. Regarding the relationship between personal characteristics and politics, BAME membership correlates with Labour Party membership to the tune of 0.124 (Spearman, p\textless0.0001) and negatively with Conservative Party membership to a lesser extent (-0.067, p\textless 0.005). BAME membership correlates positively with female gender (0.123, p\textless0.0001). Female gender correlates positively with Labour Party membership (0.240, p\textless0.0001) and negatively with Conservative Party membership (-0.199, p\textless0.0001). This group of relationships suggests women and ethnic minorities are drawn to the Labour party in an additive fashion. The correlation between BAME membership and female gender may suggest an interaction in which such cross-sectional individuals are drawn to politics, but to test this we would need a sample of non-politicians. Regarding Brexit position, Labour Party members are more likely to be remainers (0.359, p\textless0.0001) and Conservative Party members are more likely to be Leavers (0.307, p\textless0.0001); party is by far the strongest predictor of Brexit stance.

The graphs in figure~\ref{fig:sem} show, for each time period, the new, simpler SEM model. Polarity of the number on the connectors gives the regression between those two quantities. Two asterisks indicate that the relationship is significant at p\textless0.01 and one asterisk indicates that p\textless0.01. (The p-value of the overall model (given in the caption) should be interpreted slightly differently than usual when describing a SEM model; higher is better, and a satisfactory model has a p-value of greater than 0.05.)

 In summary, the main findings are:

\begin{itemize}
    \item Consistently strong relationship, predictably, between being high profile (in terms of being searched on) and receiving a lot of tweets;
    \item MPs who receive more tweets receive more abuse;
    \item But having abstracted that out, there is a variable relationship between being high profile and receiving abuse. Being famous \textit{per se} doesn't draw abuse;
    \item Engaging on Twitter consistently leads to receiving more tweets;
    \item Being female consistently leads to receiving less abuse.
\end{itemize}

%%%%%%%%%%%%%%%%%%%%%%%%%%%%%%%%%%%%%%%%%%%%%%%%%%%%

\subsection{Abuse over Time}

Harold Wilson once famously said that ``a week is a long time in politics'' and this seems especially true for the recent Brexit discussions, where a lot can change even within a single day. As such looking at month long periods may well hide interesting variations in the level of abuse MPs received. To this end, for each month we've also calculated the level of abuse received by all MPs each day and plotted that as four line graphs. It is clear from these graphs that the level of abuse MPs receive fluctuates daily, and it is not always clear why this might be. On some occasions though there are very suggestive links to important events.

\begin{figure}[t]
\begin{centering}
\includegraphics[width=\columnwidth]{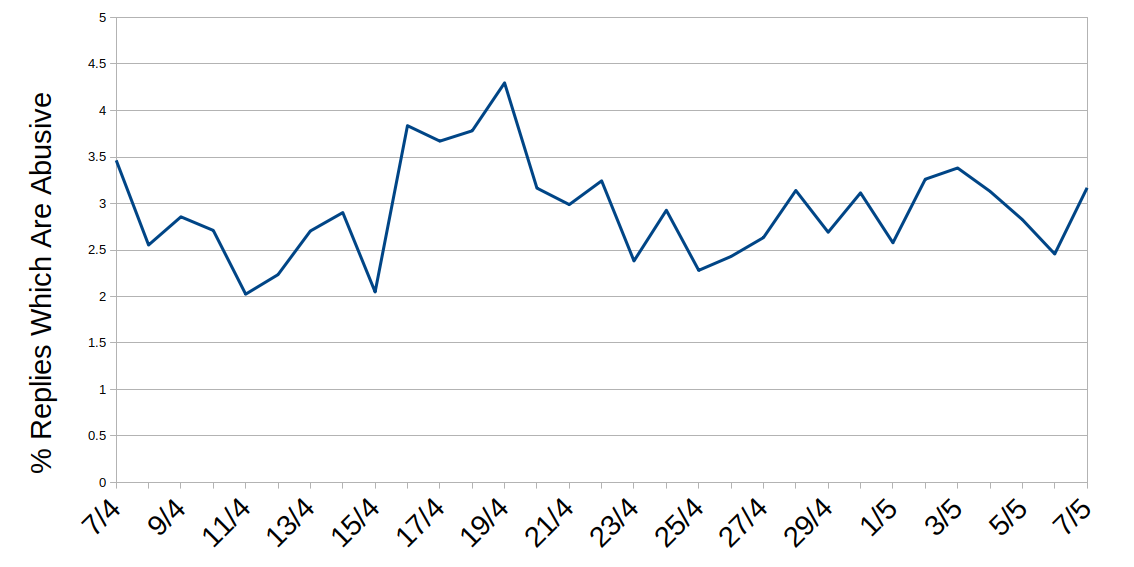}
\caption{Abuse Timeline 2015}
\label{fig:timeline2015}
\end{centering}
\end{figure}

\begin{figure}[t]
\begin{centering}
\includegraphics[width=\columnwidth]{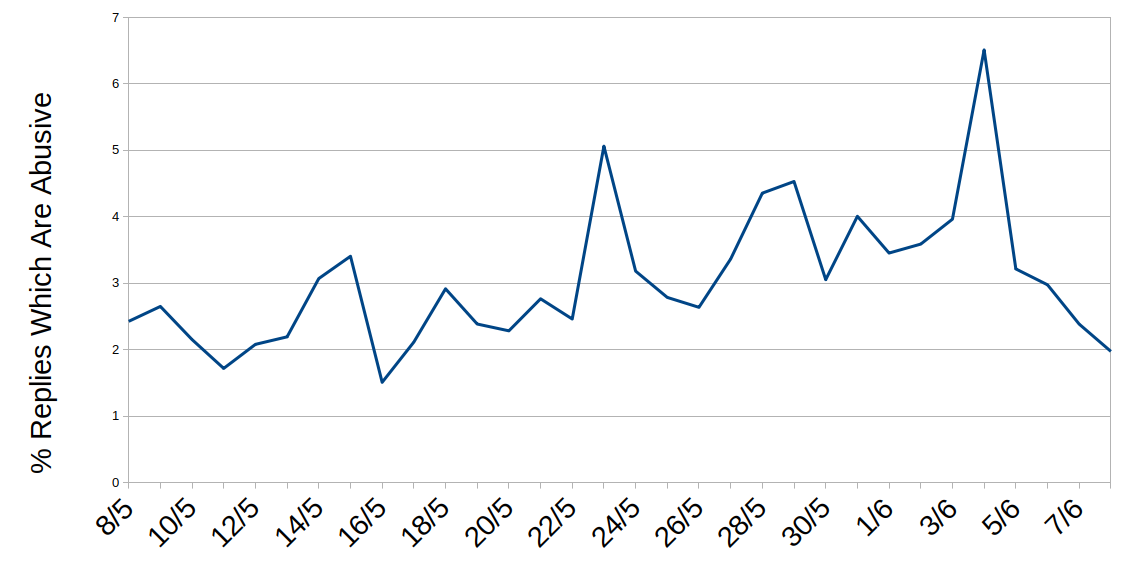}
\caption{Abuse Timeline 2017}
\label{fig:timeline2017}
\end{centering}
\end{figure}

In 2017, shown in figure~\ref{fig:timeline2017}, the two tallest spikes in abuse level focus around firstly the Manchester Arena bombing on the evening of the 22nd of May (the abuse peaks on the following day) and again around the 3rd and 4th of June after the London Bridge attack.

\begin{figure}[t]
\begin{centering}
\includegraphics[width=\columnwidth]{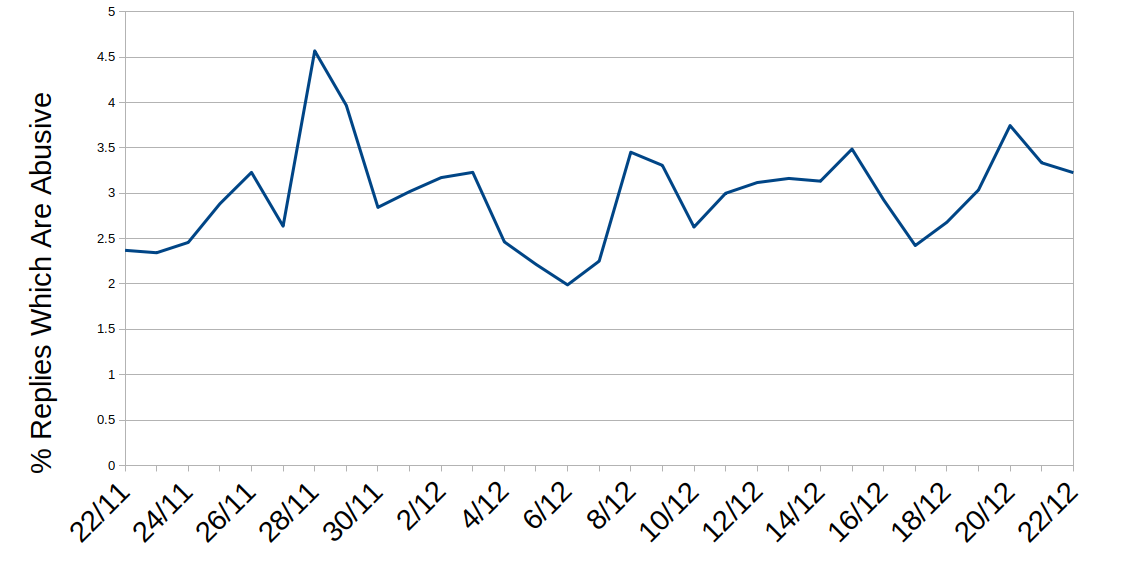}
\caption{Abuse Timeline 2018}
\label{fig:timeline2018}
\end{centering}
\end{figure}

Figure~\ref{fig:timeline2018} gives the late 2018 period. On the 25th of November 2018 the EU endorses the Brexit withdrawal agreement and abuse aimed at British MPs immediately starts to climb reaching a peak a few days later on the 28th of November which saw the first PMQs after the agreement was announced.

\begin{figure}[t]
\begin{centering}
\includegraphics[width=\columnwidth]{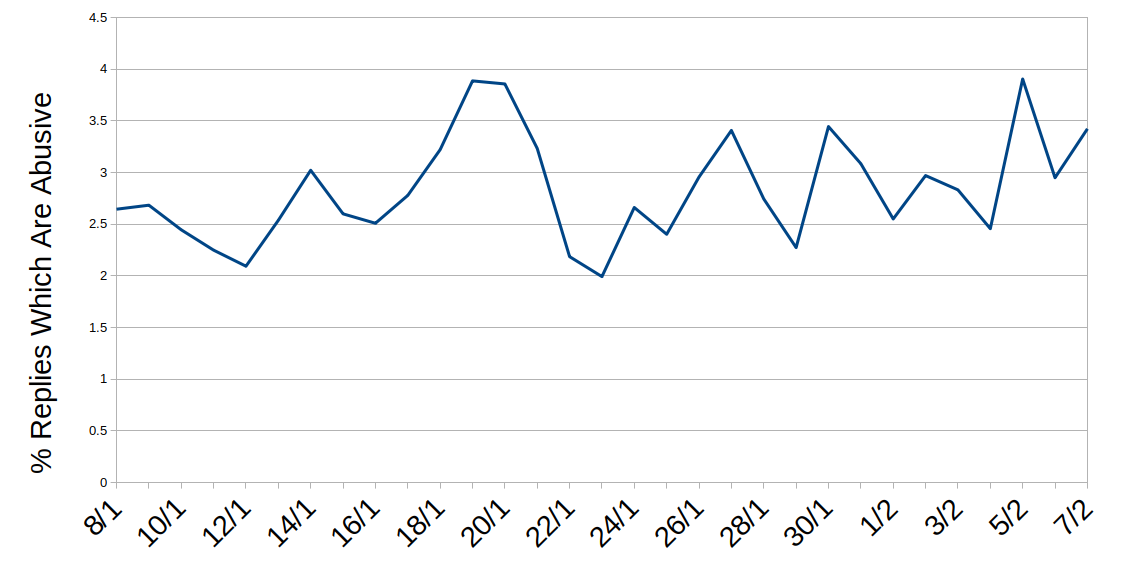}
\caption{Abuse Timeline 2019}
\label{fig:timeline2019}
\end{centering}
\end{figure}

In the final period in 2019, shown in figure~\ref{fig:timeline2019}, the wide peak starts to build on the 16th of January, the day the government survived a no confidence vote, peaking three days later on the 19th.

%%%%%%%%%%%%%%%%%%%%%%%%%%%%%%%%%%%%%%%%%%%%%%%%%%%%

\subsection{Topics which Attract Abuse}

We now turn to the topics motivating abusive tweets to politicians, across the four time periods. Abusive tweets were compared against the set of predetermined topics described earlier. Words relating to these topics were then used to determine the level of interest in these topics among the abusive tweets, in order to gain an idea of what the abusive tweeters were concerned about. So for
example, the following paraphrased 2017 tweet is counted as one for ``borders and immigration'' and one for ``schools'': \textit{``Mass immigration is
  ruining schools, you dick. We can't afford the interpretation
  bill.''}
%https://twitter.com/snotty_dog/status/872091267542470656, now suspended

We consider abuse related to specific topics both within the dataset as a whole, as well as in relation to what topics lead to abusive replies to MPs. Not every tweet contains a recognised topic, so for this analysis we are only using the tweets that do have a topic mentioned in them. The topic titles used in this section are generally self-explanatory, but a few require clarification. ``Community and society'' refers to issues pertaining to minorities and inclusion, and includes religious groups and different sexual identities. ``Democracy'' includes references to the workings of political power, such as ``eurocrats''. ``National security'' mainly refers to terrorism, where ``crime and policing'' does not include terrorism. The topic of ``public health'' in the UK is dominated by the National Health Service (NHS). ``Welfare'' is about entitlement to financial relief such as disability allowance and unemployment cover.

\subsubsection{Topics in the whole dataset}

This subsection is the first time in this work that we make use of the entire dataset, rather than just tweets to sitting MPs. Table~\ref{tab:topics-dataset} contains statistics for this corpus. You can see the percentage of abusive tweets in each of the time periods. This is notably lower than for the tweets only to MPs.

\begin{table}
\begin{center}
\resizebox{.7\columnwidth}{!}{%
  \begin{tabular}{l|l|l|l}
  \textbf{Date} & \textbf{Tot.} & \textbf{Ab.} & \textbf{\% Ab.}\\
  \hline
2015 & 3,136,352 & 19,094 & 0.609\\
2017 & 4,639,050 & 39,078 & 0.842\\
2018 & 4,738,168 & 60,110 & 1.269\\
2019 & 4,561,468 & 55,372 & 1.214\\
 \end{tabular}
 }
\caption{Statistics for entire dataset}
\label{tab:topics-dataset}
\end{center}
\end{table}

In this analysis we use only the tweets that mention a topic. In table~\ref{tab:topic-subset} you can see how many that is--roughly one third to a half of tweets mention at least one of the selected topics. Total numbers of original tweets, replies and retweets are given. In this section we use, at different points, \textit{all retweets that have a topic mention}, and \textit{abusive replies to all politicians in the dataset} (not just sitting MPs but also candidates, and MPs from other time periods). For this reason, in the table below, in addition to breaking down the total number into original tweets, replies and retweets, we also include information about the number of replies to politicians that are abusive. These are relatively small in number.

\begin{table*}
\begin{center}
\resizebox{.7\textwidth}{!}{%
  \begin{tabular}{l|l|l|l|l|l|l}
  \textbf{Date} & \textbf{All tweets} & \textbf{Original tweets} & \multicolumn{3}{l}{\textbf{Reply tweets to politicians}} & \textbf{Retweets}\\
  \hline
   & \textbf{Total} & \textbf{Total} & \textbf{Total} & \textbf{Abusive} & \textbf{\% Abusive} & \textbf{Total}\\
  \hline
2015 & 1,013,679 & 63,329 & 192,906 & 2,426 & 1.258 & 757,444\\
2017 & 1,819,657 & 44,286 & 296,233 & 6,562 & 2.215 & 1,479,138\\
2018 & 2,590,070 & 40,661 & 466,240 & 11,242 & 2.411 & 2,083,169\\
2019 & 2,507,191 & 39,218 & 475,344 & 11,112 & 2.338 & 1,992,629\\
 \end{tabular}
 }
\caption{Subset of entire dataset that contains a topic mention}
\label{tab:topic-subset}
\end{center}
\end{table*}

Next, we examine which topics featured most in abusive replies to politicians over the respective time periods and compare these to the topics that elicited engagement through retweets.

\begin{figure}
  \centering
    \includegraphics[width=\columnwidth]{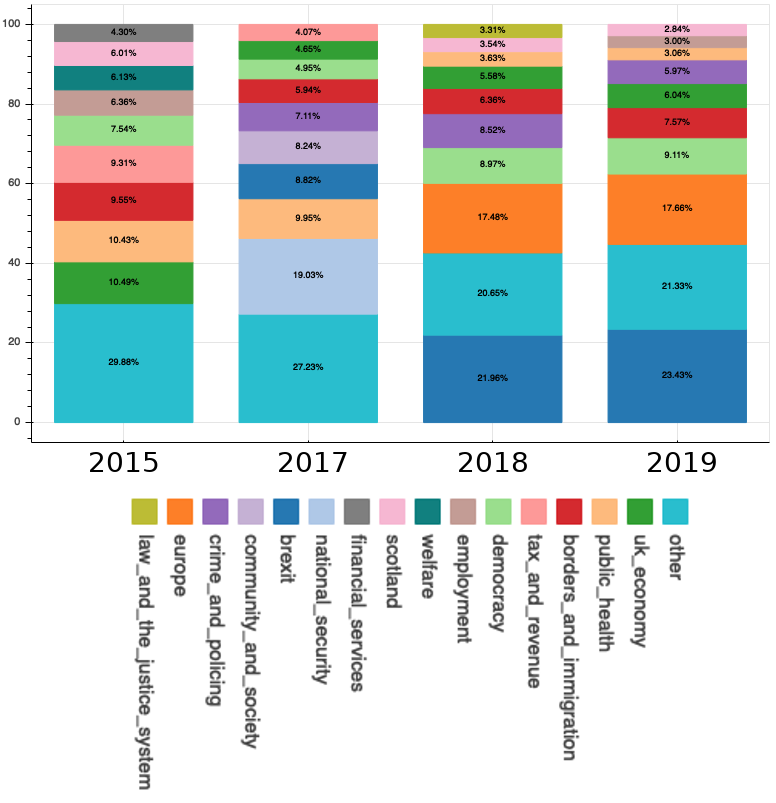}
  \caption{Topics in Abusive Replies to Politicians}
\label{fig:topics-pol}
\end{figure}

\begin{figure}
  \centering
    \includegraphics[width=\columnwidth]{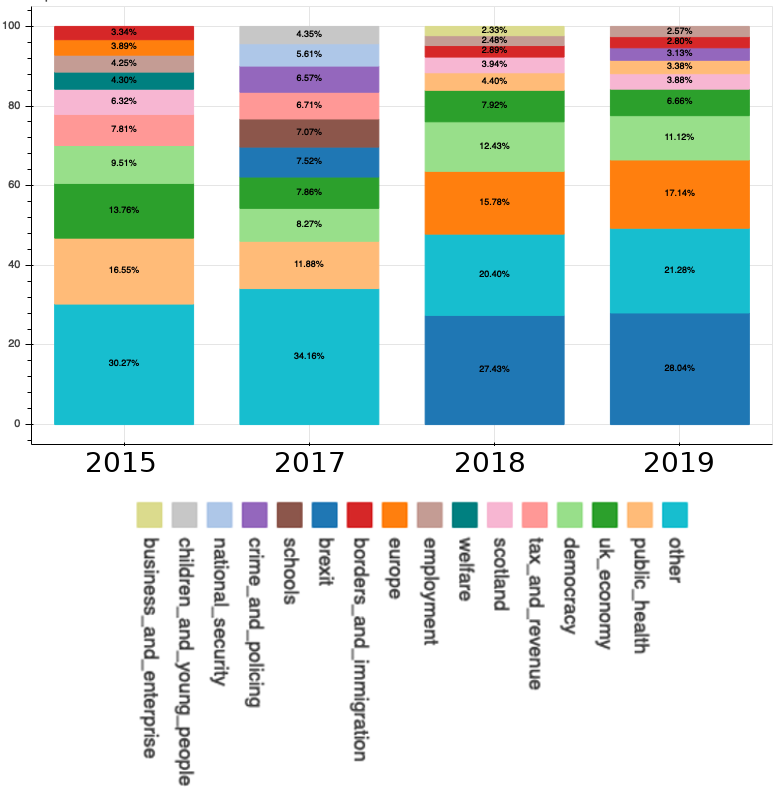}
  \caption{Topics in All Retweets containing a Topic Mention}
\label{fig:topics-all-retw}
\end{figure}

What the stacked bar charts in figures~\ref{fig:topics-pol} and \ref{fig:topics-all-retw} clearly show is the rise of Brexit and Europe as topics that trigger online abuse and political debate. In 2015 Brexit did not appear within the top 9 topics, as it was well before the referendum date had even been set. Likewise, the related topic of Europe, which concerns EU, EC, and MEPs, attracted less than 4\% of retweets and did not feature amongst the top 9 topics in abusive replies. By the 2019 period, however, over a quarter of all retweets with a topic are about Brexit and over 17\% are about Europe, with the same trend observed in abusive replies. This clearly demonstrates how Brexit and Europe have come to dominate political discourse, as the primary topics of engagement with MPs, polarised debates and online abuse.

It is notable that in 2017 national security comes to the fore in abuse-containing tweets, whilst only being the ninth most prominent topic in the retweets. Similarly, community and society is more frequent in abuse-containing tweets than in the retweets. In the month preceding the 2017 election, the UK witnessed its two deadliest terrorist attacks of the decade so far, both attributed to ISIS. In 2015 economy is the most prominent topic in abusive tweets and a very close second in the retweets. National security was not an important topic in 2015. However, borders and immigration  appears  more  prominently  in  the  abuse-containing tweets.

\begin{figure}
  \centering
    \includegraphics[width=.95\columnwidth]{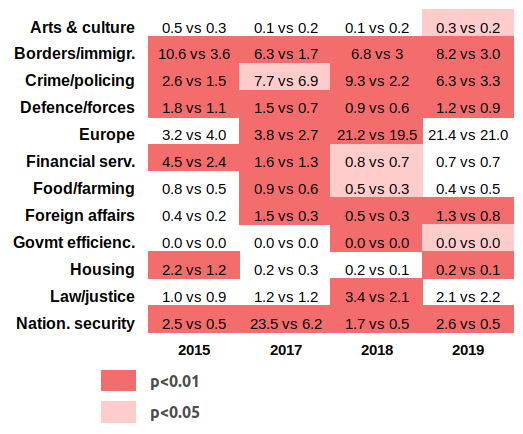}
  \caption{Topics attracting more abuse than is typical (\%)}
\label{fig:topics-more}
\end{figure}

\begin{figure}
  \centering
    \includegraphics[width=.95\columnwidth]{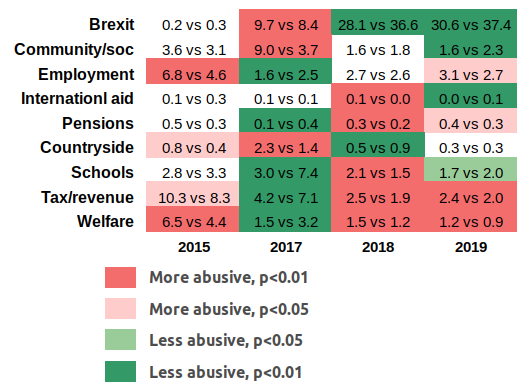}
  \caption{Topics that vary in abuse attracted (\%)}
\label{fig:topics-var}
\end{figure}

\begin{figure}
  \centering
    \includegraphics[width=.95\columnwidth]{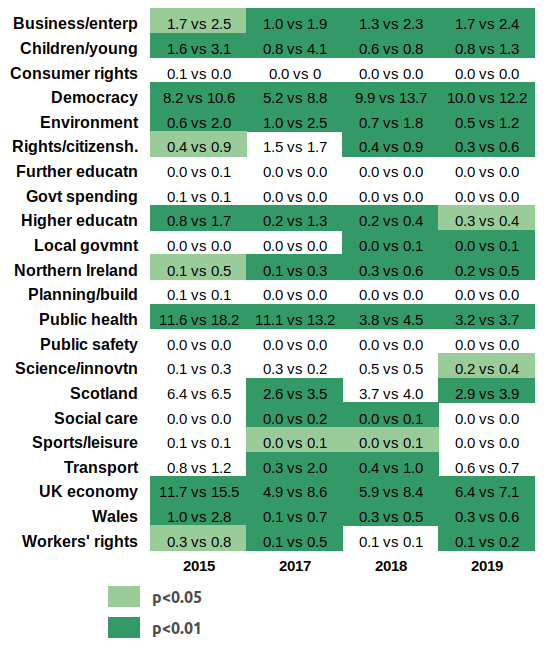}
  \caption{Topics attracting less abuse than is typical (\%)}
\label{fig:topics-less}
\end{figure}

The heat map charts in figures~\ref{fig:topics-more}, \ref{fig:topics-var} and \ref{fig:topics-less} illustrate which topics attracted more abuse than is typical for the tweets from that period, or less abuse than is typical, for each time period. This is done by comparing the proportion of tweets on a particular topic that are abusive with the proportion of the entire dataset that is abusive, using Fisher's exact test. A dark red block indicates that that topic in that year attracted very significantly more abuse than the background distribution (Fisher's exact test, p\textless0.01). Light red indicates a result significant at p\textless0.05. Light green indicates that that topic attracts less abuse than average for the corpus (p\textless0.05) and dark green indicates that that topic in that time period attracted very significantly less abuse (p\textless0.01).
Note that the significance of the difference between incivility in tweets on the topic and the background, shown using colour in the heat map, does not indicate the magnitude of the difference. Numbers in the cells therefore give firstly the percentage of topic mentions in abusive tweets, and secondly the percentage of topic mentions in non-abusive tweets.

We can see that some topics have consistently attracted abuse (table~\ref{fig:topics-more}): borders and immigration; crime; defence; and national security (terrorism). Others have tended to invite a civil tone, or at least fail to invite an abusive one (table~\ref{fig:topics-less}): children; the environment; the economy; business; Northern Ireland; and Wales. Topics with no significant differences from the background distribution have also been included in this table.

Several topics show variation and possibly an evolution over time in the civility of discussion attracted (table~\ref{fig:topics-var}). In particular, we note that the Brexit discussion in 2017 was uncivil, but by the 2018 and 2019 periods it had become civil. There is a suggestion of a ``2017 effect'', in that often it is the 2017 time period that shows a different response to that topic than usual. In 2017:

\begin{itemize}
    \item ``Brexit'', ``community and society'' and ``rural and countryside'' attracted unusually abusive attention;
    \item ``Employment'', ``pensions and ageing society'', ``tax and revenue'' and ``welfare'' attracted unusually civil attention.
\end{itemize}

It is possible that issues that normally inspire impassioned feeling were displaced from the public consciousness in the wake of the referendum, when Brexit itself and matters of community and belonging were brought to the fore. ``Rural and countryside'' issues are highly relevant in the context of Brexit. ``International aid and development'' and ``schools'' don't fit a 2017 pattern.

\subsubsection{Topics which Drive Abusive Replies}

The above analysis focuses purely on the topics of the retweets and abusive replies to politicians across the entire dataset. Next, we examine variations in abuse topics at a party level. The charts below focus only on the three main parties as they receive large enough volumes of abusive replies to give meaningful results.

This analysis differs from that in the previous section as it looks at the original tweets written by MPs in conjunction with the replies they receive. As with previous analysis we start by collecting all the abusive replies sent to MPs, but we use them only as a stepping stone to find the tweets an MP sent to which they were in reply. These original MP tweets are then processed to determine which topics they contain. The idea is to see if tweets on different topics results in the differing levels of abusive replies.

We present this analysis in the following four graphs (one per time period) in which we display the top nine topics that triggered abusive replies, and the remaining tweets are grouped under other (this is done not in an attempt to hide the other topics but simply to make the graphs readable). For example, looking at the graph for 2015 we can see that just under 20\% of the abusive replies to Conservative MPs were triggered by the MP tweeting about employment, whereas just over 20\% of the replies to Labour were in response to tweets by their MPs on the topic of welfare.

\begin{figure}
  \caption{Topics in abuse in replies to politicians of different parties: 2015}
  \centering
    \includegraphics[width=\columnwidth]{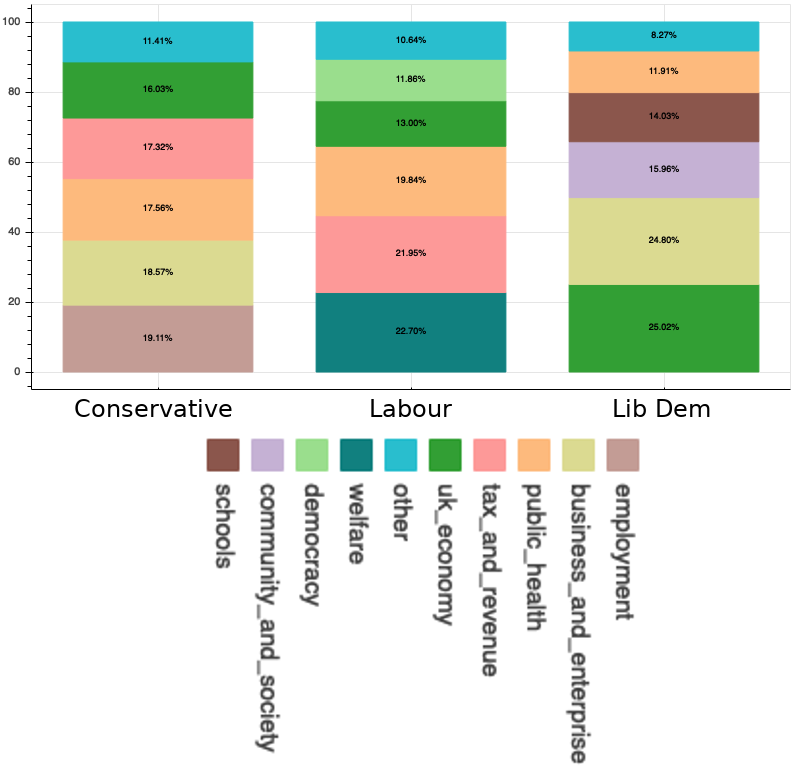}
\label{fig:abuse-party-2015}
\end{figure}

\begin{figure}
  \caption{Topics in abuse in replies to politicians of different parties: 2017}
  \centering
    \includegraphics[width=\columnwidth]{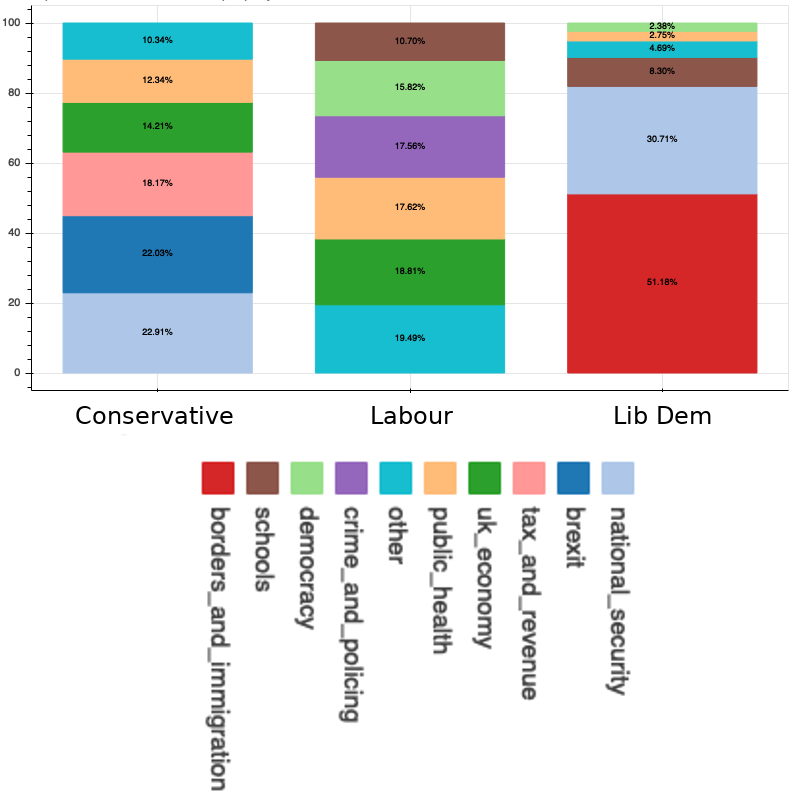}
\label{fig:abuse-party-2017}
\end{figure}

\begin{figure}
  \caption{Topics in abuse in replies to politicians of different parties: 2018}
  \centering
    \includegraphics[width=\columnwidth]{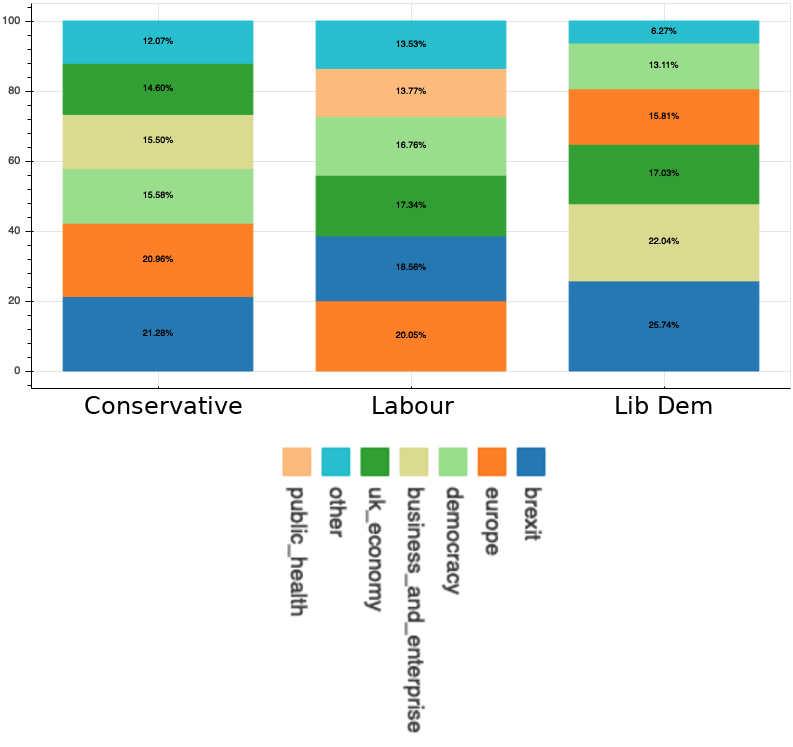}
\label{fig:abuse-party-2018}
\end{figure}

\begin{figure}
  \caption{Topics in abuse in replies to politicians of different parties: 2019}
  \centering
    \includegraphics[width=\columnwidth]{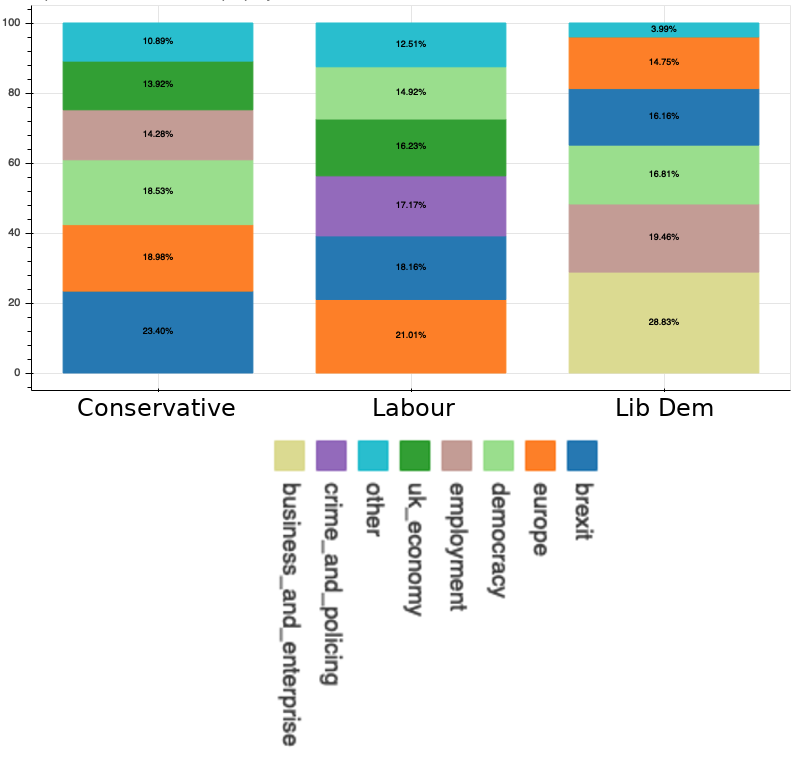}
\label{fig:abuse-party-2019}
\end{figure}

Again these graphs show that in 2018 and 2019 for the two main parties (Conservatives and Labour) Brexit and the wider topic of Europe are the primary focus of abusive replies. In comparison, in 2015 the topic of Europe was not amongst the most prominent topics of abuse (employment, business, economy, public health, etc), whereas in 2017 Brexit appears as a topic only for the conservative party.

Subjects attracting ire differ between parties, and may give an indication of how the party (or its key personnel) are viewed. Employment has arisen in the top topics for the Conservative party twice, but not Labour. Crime and policing has arisen twice in top topics for the Labour party, but no other. It is striking that tweets about borders and immigration drew abusive responses for the Liberal Democrats in 2017, but don't appear in top topics for any other party or time period, given that borders and immigration is a prominent topic in abusive tweets. Similarly, tweets about national security drew more abusive replies proportionally for the Liberal Democrats in 2017 than for the Conservatives, whilst national security doesn't feature among top topics for Labour. The tone of Liberal Democrat tweeting regarding immigration and terrorism was unambiguously left-wing, whereas Labour tended to focus on criticizing the Conservative Party whilst hedging about their own position.

\subsection{How Abusive Words are Used}

In this section we present some exploration into how words are used. We start with work on how women and men are differently addressed, before turning to trends in word use over time.

\subsubsection{Words used differently towards male and female MPs}

In table~\ref{tab:gendered-abuse}, Fisher's exact test was used to establish if a word is significantly more often used in a tweet to a female or a male MP, as a proportion of all abuse terms found in tweets to that group. (This may occasionally mean that a word appears on the female list because it is used unexpectedly frequently given that women receive less abuse, even though it may be similarly or more frequently used against male MPs on a per-individual basis.) Only abuse terms that appear more than 500 times across all the years/genders combined are included in the table. Stronger words, selected according to the Ofcom guide\footnote{\small{\url{https://www.ofcom.org.uk/__data/assets/pdf_file/0022/91624/OfcomOffensiveLanguage.pdf}}} or in keeping with it, are indicated in bold.

\begin{table}
\begin{center}
\resizebox{\columnwidth}{!}{%
  \begin{tabular}{l|l|l|l}
\textbf{Word group} & \textbf{Levelled More} & \textbf{Levelled More} & \textbf{No}\\
& \textbf{at Female MPs than} & \textbf{at Male MPs than} & \textbf{significant}\\
& \textbf{Male (p\textless0.01)} & \textbf{Female (p\textless0.01)} & \textbf{difference}\\
\hline
\hline
Gendered & \textbf{'witch'}, \textbf{'slag'}, & \textbf{'wanker'}, \textbf{'bastard'}, & \\
 & 'bitch' & \textbf{'tosser'}, \textbf{'dickhead'} & \\
 \hline
Fem bdy pt &  & \textbf{'cunt'}, \textbf{'twat'}, 'tit'
& \textbf{'twats'}\\
as insult & & &\\
\hline
Male bdy pt &  & \textbf{'prick'}, \textbf{'dick'},&\\
as insult &  & \textbf{'bellend'}, \textbf{'plonker'} &\\
\hline
Reference to & 'loony'  & 'muppet', 'you idiot', & 'cretin', 'idiot',\\
intelligence &  & \textbf{'fuckwit'}, 'pillock', & \textbf{'moron'}\\
 &  & 'numpty' & \\
\hline
Ref. to dirt & \textbf{'scum'}, \textbf{'turd'} & & 'scumbag'\\
\hline
Aggression & \textbf{'fuck off'}, \textbf{'fuck you'} & & 'stfu'\\
\hline
Other & 'coward', & 'git', 'prat', & 'pish'\\
 & 'your arse' (p\textless0.05) & 'arsehole' & \\
 \end{tabular}
 }
\caption{Abuse Words Used Differently Toward Women and Men MPs}
\label{tab:gendered-abuse}
\end{center}
\end{table}

The impression given by the table reinforces the above finding that men tend to receive more abuse, with the suggestion of a richer vocabulary of insults being targeted to a greater extent at men. Even terms that need not technically be considered gendered, such as ``bastard'' and ``git'', still suggest a male target through routine usage, making them hard to classify. It is notable however that women are more likely to be addressed with ``fuck off'', or ``fuck you'' (as a proportion of all abuse words received), phrases which are perhaps unified by being expressions of aggression, and by the words ``scum'' and ``turd'', which might be thought of as ``dirt'' words.

\begin{figure}
  \centering
    \includegraphics[width=\columnwidth]{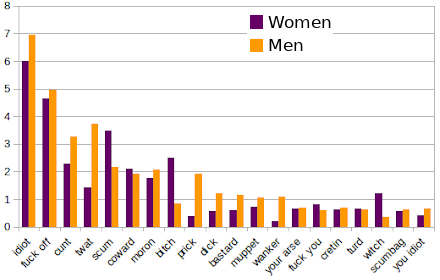}
  \caption{Top 20 Abuse Words, Instances per MP}
\label{fig:top-abuse-words}
\end{figure}

The bar chart in figure~\ref{fig:top-abuse-words} gives an indication in quantitative terms of which abuse words are being used in tweets to male and female MPs (the table above gives significant differences but not an idea of the extent of word use or how male-female experience differs in absolute terms). The y-axis gives counts of this term being used per person, averaged over the four time periods, for the 20 most used abuse terms. So, for example, on average a woman MP might expect to be addressed with ``idiot'' six times in any one time period (though note that abuse does not spread itself in a normal distribution, so ``average'' is not a usual case, most MPs receiving less than that, and a few much more).

\subsubsection{Strength of language used against male and female MPs}

Words were again classified according to Ofcom's guide to degree of offensiveness.\footnote{\small{\url{https://www.ofcom.org.uk/__data/assets/pdf_file/0022/91624/OfcomOffensiveLanguage.pdf}}} Where a word was not found in the guide, it was classified in a way that was in keeping with the guide, if this could be cautiously done (for example ``fuckwit'' was considered "strongest" because Ofcom positions ``fuck'' in the strongest category). Spelling variations were also considered, as well as pluralization. Words not included in Ofcom or very similar to words included were excluded from this analysis. Words were then grouped into two categories; mild/medium and strong/strongest, and male and female scores compared for the two groups. Where words were classified into binary male/female categories according to whether men or women were more targeted by that word as a percentage of total abuse received by that group, women are significantly more likely to be targeted with milder words (t-test, p\textless0.05). Where the test was performed on the absolute difference between the words as a percentage of total abuse received by men and women, again women were significantly more likely to be targeted with milder words (t-test, p\textless0.05). Where the difference was calculated between the average number of times a woman was targeted with that word in contrast to a man, the pattern still holds but the statistical significance is weaker (p\textless0.1). In summary, there is some evidence that milder words are used against women compared with men.

\subsubsection{Word usage change across time}

\begin{figure}
  \centering
    \includegraphics[width=\columnwidth]{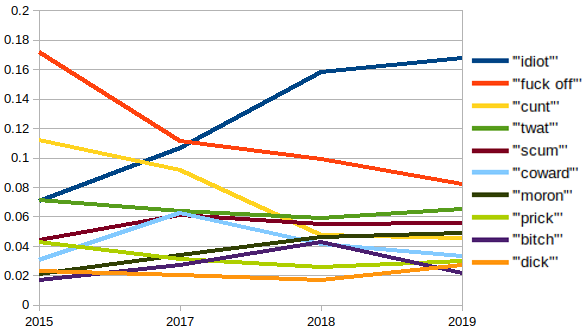}
  \caption{Abuse Words over Time (Proportion of Total for Year)}
\label{fig:abuse-words-time}
\end{figure}

Over time, there is more change in abuse term usage between 2015 and 2017, and between 2017 and 2018. Between 2018 and 2019 periods little time elapsed, and consequently most changes in abuse term usage are not significant. The ten most common abuse words across the four time periods are plotted in figure~\ref{fig:abuse-words-time}, in terms of the proportion of abuse words in that time period (e.g. 17\% of the abuse in 2015 was ``fuck off'').

There is a suggestion of a trend over time towards milder abuse words. Note for example that ``fuck off'' has declined, whilst ``idiot'' has risen. In order to test this, words were again grouped according to Ofcom's guide to degree of offensiveness into two categories; mild/medium and strong/strongest. The rise and fall of the terms was calculated in terms of the change in proportion of total abuse words used in that year between time periods (for example, ``your arse'' declined from 1.4\% in 2015 to 1\% in 2017, giving -0.4\%). These change figures were then compared for the two word groups using a t-test. Generally speaking, milder words rose and stronger words declined, except for between 2018 and 2019, where the time elapsed was short. The difference between groups was statistically significant between 2017 and 2018 (p\textless0.05). Whilst not statistically significant, the change between 2015 and 2017 and the overall change between 2015 and 2019 was in keeping with the trend described.

It may be that awareness has increased and/or attitudes have changed, or it may be that users hope to avoid sanctions from Twitter, resulting in an increase in use of milder words such as "idiot" and a decrease in stronger words/phrases such as ``fuck off'' and ``cunt'', as illustrated in the chart above.

%%%%%%%%%%%%%%%%%%%%%%%%%%%%%%%%%%%%%%%%%%%%%%%%%%%%%%%%%%%%%%%%%%%%%%%%

\section{Discussion and Future Work}

We describe how abusive language was used in tweets to British MPs in four separate month-long time periods; preceding general elections in 2015 and 2017, and around key decision points in the negotiation of the United Kingdom's exit from the European Union late 2018 and early 2019. The data show what politicians contend with in terms of abusive attention on Twitter, and how variable this between individuals with different public roles and reputations.

We find that the topics that draw abuse evolve over the period studied. We see the appearance of ``Brexit'' as the main focus of attention both in abusive tweets and generally, drawing attention from the previously strong focus of public health (the NHS). Among abusive tweets, Brexit has captured attention previously focused to a greater extent on borders and immigration. In the wake of the referendum, national security (terrorism) drew abusive attention in a way not seen in any other period or in a non-abusive tweets.

The 2017 period preceded a general election uniquely focused on Brexit, as Theresa May sought to strengthen her negotiation position. In various ways, the focus of attention in that time period differed from the others. The topic of Brexit itself attracted unusually abusive attention, as did ``community and society'', a topic that includes, most relevantly, religious identities. After that period, the tone of conversation about Brexit and about identity groups became civil. Recall that there were two terrorist attacks attributed to ISIS in the UK in the month preceding the 2017 general election, whilst the threat level from ISIS has dropped considerably since.

With regards to the relationship between demographic characteristics and abuse, much depends on individual characteristics such as an individual's being outspoken on certain topics. Being prominent \textit{per se} doesn't particularly attract abuse, but more prominent individuals will draw more attention on Twitter, and therefore more abuse by volume. Whilst Conservatives seem to attract more abuse, this effect is lost when prominence is factored out. However, female gender does seem to lead to less abuse by proportion in quantitative terms. There is also some evidence that women are subjected to less strong abuse. However, they were more likely to be addressed using the ``f'' word and words suggesting dirt.

The finding regarding women receiving less abuse is in keeping with findings across a more general population~\cite{pew2014}, though Pew find women in the 18-24 age range are disproportionately targeted, whilst there are no women of that age range in our sample. The reader should also note that words mean different things depending on who is addressed, and quantitative findings do not address the full complexity of society. Abuse might be received very differently, for example, by a person who is more physically or socially vulnerable, or where it attacks a person's right to take their place in the political arena. Furthermore our approach lumps abuse together. It may be that a general tendency to address women somewhat more politely masks a sinister minority of more unpleasant abuse. Most of the abuse in our sample arises from individuals expressing a strength of feeling about political matters, which isn't personal. We don't present data here about the specific demographics of personal abuse, threats, bullying or persecution.

%%%%%%%%%%%%%%%%%%%%%%%%%%%%%%%%%%%%%%%%%%%%%%%%%%%%%%%%%%%%%%%%%%%%%%%%

\section{Acknowledgments}
\small{
This work was partially supported by the European Union under grant
agreements No. 654024 SoBigData and No. 825297 WeVerify.

\bibliographystyle{aaai}
%\bibliography{../../../../../../sale/big}
\bibliography{politicians-twitter-abuse-arxiv}
}
\end{document}